\documentclass[letterpaper,12pt]{JHEP3}
\usepackage{graphicx} 
\usepackage{amsmath} 
\usepackage{amsfonts} 
\usepackage{amssymb} 
 
\def\beq{\begin{equation}} 
\def\eeq{\end{equation}} 
\def\bea{\begin{eqnarray}} 
\def\eea{\end{eqnarray}} 
\def\ben{\begin{enumerate}} 
\def\een{\end{enumerate}}

\def\lsim{\mathrel{\raise.3ex\hbox{$<$\kern-.75em\lower1ex\hbox{$\sim$}}}} 
\def\gsim{\mathrel{\raise.3ex\hbox{$>$\kern-.75em\lower1ex\hbox{$\sim$}}}} 
\def\ifmath#1{\relax\ifmmode #1\else $#1$\fi}

\def\gev{~{\mbox{GeV}}} 
\def\GeV{~{\mbox{GeV}}}

\def\cnone{\wt\chi^0_1} 
 
\def\cntwo{\wt\chi^0_2}

\def\se{\wt e} 
\def\smu{\wt\mu}

\def\gr{\wt G}

\def\wt{\widetilde}

\def\cpone{\wt \chi^+_1} 
\def\cmone{\wt \chi^-_1} 
\def\cpmone{\wt \chi^{\pm}_1}

\def\stau{\wt \tau}

\def\sbot{\wt b}

\def\slep{\wt \ell}

\newcommand{ \slashchar }[1]{\setbox0=\hbox{$#1$}   
   \dimen0=\wd0                                     
   \setbox1=\hbox{/} \dimen1=\wd1                   
   \ifdim\dimen0>\dimen1                            
      \rlap{\hbox to \dimen0{\hfil/\hfil}}          
      #1                                            
   \else                                            
      \rlap{\hbox to \dimen1{\hfil$#1$\hfil}}       
      /                                             
   \fi}

\title{Missing Momentum Reconstruction and Spin Measurements at Hadron Colliders} 
\author{Hsin-Chia Cheng${}^{a}$, Zhenyu Han${}^{b}$,  Ian-Woo Kim${}^{c}$ and Lian-Tao Wang${}^{d}$
\\{ \small \sl ${}^{a}$Department of Physics, University of 
  California, Davis, CA 95616\\  
${}^{b}$Department of Physics, Harvard University, Cambridge, MA 02138\\
${}^{c}$Department of Physics, University of Wisconsin, Madison, WI 53706\\
${}^{d}$Department of Physics, Princeton University, Princeton, NJ 08544} } 
\abstract{  
We study methods for reconstructing the momenta of invisible particles in cascade decay chains at hadron colliders. We focus on scenarios, such as SUSY and UED,  in which new physics particles are pair produced. Their subsequent decays lead to two decay chains ending with neutral stable particles escaping detection. Assuming that the masses of the decaying particles are already measured, we obtain the momenta by imposing the mass-shell constraints. Using this information, we develop techniques of determining spins of particles in theories beyond the standard model. Unlike the methods relying on Lorentz invariant variables, this method can be used to determine the spin of the particle which initiates the decay chain. We present two complementary ways of applying our method by using more inclusive variables relying on kinematic information from one decay chain, as well as constructing correlation variables based on the kinematics of both decay chains in the same event. 
} 
 
\begin{document} 
\section{Introduction} 
\label{sec:introduction} 

The operation of the Large Hadron Collider (LHC) starts a new era in high energy physics of direct exploration into the TeV scale. 
New physics beyond the Standard Model (SM) is strongly expected to occur at the TeV scale because of the hierarchy problem. 
Another strong hint for new physics at the TeV scale comes from the dark matter (DM) in the universe.
It is now well-established that $\sim 23\%$ of the total energy of the whole universe is made of dark matter,
 and it cannot be accounted for from any SM particles. 
The leading candidate for the dark matter is a new weakly interacting massive particle (WIMP) with a mass in the range of $\sim 10$ GeV to a few TeV\footnote{See Ref.~\cite{Feng:2010gw} for a recent review.}. The thermal relic of such a particle from the Big Bang can give the right amount of dark matter in the universe if its interactions with SM particles and itself are of the similar strength of the weak interaction. To be stable it should be charged under a new symmetry ({\it e.g.,}  a $Z_2$ parity as the simplest example).

Many scenarios of TeV new physics beyond the Standard Model have been proposed to address the hierarchy problem. They often also contain  a dark matter candidate, {\it e.g.,} supersymmetry (SUSY) with $R$-parity, Universal Extra Dimensions (UEDs) with Kaluza Klein (KK) parity~\cite{Appelquist:2000nn,Cheng:2002iz,Cheng:2002ab}, little Higgs models with $T$-parity~\cite{Cheng:2003ju,Cheng:2004yc}, warped extra dimensions with a $Z_3$ symmetry~\cite{Z3} and so on. A common feature of these models is that there are other new particles charged under the same new symmetry which protects the stability of the DM particle. These new particles may be pair produced copiously at the LHC. After production they will go through cascade decays to the lightest one which escapes the detector. Therefore, many of these different models can give rise to similar collider signatures, {\it i.e.,} jets/leptons with missing transverse momentum. It is important to be able to distinguish different models if such experimental signals are found. In particular, the spin measurements are essential to distinguish SUSY where the spins of the SM particles and their superpartners differ by 1/2, from other models where the partners have the same spins as the corresponding SM particles.

To determine the spin of a particle in a more model-independent way, we need to examine the angular distributions of its production or decay.
Although theoretically well-motivated, these collider signatures with missing transverse momentum pose a serious challenge to such experimental measurements at hadron colliders. Because the new particles are pair-produced, there are at least two missing particles (one from each chain) in every event. For any given single event there is not enough information to reconstruct the full kinematics without additional information. 
 In a long decay chain, the polar angle of the decay of the intermediate particle in its rest frame is directly related to the Lorentz invariant mass combination of the visible particles of the decay chain. One can use the invariant mass distribution to determine the spin of the intermediate particle without fully reconstructing the kinematics if certain conditions are satisfied. Many of the spin determination methods in the literature are based on this observation~\cite{Barr:2004ze, Smillie:2005ar, Datta:2005zs,  Alves:2006df, Athanasiou:2006ef, Wang:2006hk, Smillie:2006cd, Kilic:2007zk, Csaki:2007xm,Wang:2008sw, Burns:2008cp, Gedalia:2009ym, Ehrenfeld:2009rt, Kong:2010mh}. However, it cannot be used to completely determine the spins of the first and the last particles in a decay chain directly. If the momenta of the invisible particles of each event can be reconstructed, then one can boost the event to any frame and examine any relevant kinematic distributions. The spin of a particle can be determined from the azimuthal-angle correlations~\cite{azimuthal} as well as the polar angle distributions. In particular, one can determine the spin of the first particle in a decay chain by looking at the angular distributions of its production or decay.

It is possible to reconstruct the momenta of the invisible particles if there are enough constraints to match the number of unknown kinematic variables in the event, {\it e.g.,} if there are enough mass shell constraints and the masses of the particles in the decay chains are already known. Measuring the masses of the particles in a decay chain with missing transverse momentum itself is a non-trivial task, as there is no invariant mass peak and the visible momenta are more sensitive to the mass differences than the absolute masses. There have been many research efforts recently in mass determinations for various event topologies. A lot of progress has been made and many new methods have been proposed based on various kinematic variables and constraints (for a review, see \cite{mass_review}). We expect that the masses of the new particles can be quite accurately determined if a substantial clean signal sample can be isolated and the visible momenta are well measured. In particular, for extended decay chains we consider in this article, the masses can be determined with a few percent errors using a few hundred events \cite{Kawagoe:2004rz, Cheng:2007xv, mass33, Nojiri:2010dk}. 

The rest of this paper is organized as follows. In Sec.~\ref{sec:general}, we discuss single-chain and double-chain event topologies which can be kinematically reconstructed with the mass measurements. In Sec.~\ref{sec:single}, we discuss spin determination from a single decay chain. In particular, we identify a case where we can determine the spin of the first decaying particle, which cannot be measured by the invariant mass technique. In Sec.~\ref{sec:double}, we discuss the double-chain techniques, where additional information from the spin correlation between the two decay chains may be used. Further discussion and conclusions are drawn in Sec.~\ref{sec:conclusions}.

\section{General Considerations of Event Reconstruction}
\label{sec:general}

As we discussed in the Introduction, it is very useful if we can reconstruct the invisible particles' momenta from the visible momenta and available kinematic constraints. In this section, we give a general counting of constraints for different topologies of events with missing transverse momentum, and discuss the corresponding methods for event reconstruction. As we will see, this depends on whether we are examining a single decay chain or both decay chains in an event, and whether the system is under-constrained, exactly-solvable or over-constrained.  
  
The unknowns in the problem are the 4-momenta of the missing particles. Assuming that there is only one missing particle in each decay chain, we have 4 unknowns for each event if we want to reconstruct only one of the decay chains, and 8 unknowns if we want to reconstruct both decay chains. As we mentioned in the Introduction, we will assume that the masses of all new particles in the decay chains are already measured with some errors. Each on-shell particle then contributes a constraint on the missing momenta. These ``mass-shell constraints'' are available for both single-chain and double-chain cases. For the double-chain case, two more constraints are available from the measured missing transverse momentum if there are no extra missing particles. It is then straightforward to count the number of constraints needed for event reconstruction. For the single-chain case, we need 4 mass shell constraints to solve the system. (We say the system is {\it exactly-solvable} when the number of unknowns is equal to the number of constraints.) This corresponds to a decay chain with 3 visible particles (including particles decaying further but not introducing extra invisible particles, such as a $Z$-boson decaying to charged leptons/quarks) if all decays are two-body. If the decay chain is longer, we have an over-constrained system and we can employ a likelihood method to obtain the best-fit missing momenta. If the decay chain is shorter, the missing momenta cannot be fully reconstructed. Similarly, in the double-chain case, we need 6 mass-shell constraints, which, together with the constraints from the measured missing transverse momentum, allow us to solve the system. An example is the case with two on-shell decays for each decay chain. This occurs for $t\bar t$ pair production in the dilepton decay channel and can be used for determining the spin of a $t\bar t$ resonance \cite{ttbar}. Again, if the decay chains are longer (shorter), we have an over-(under-)constrained system. 

The single-chain case and double-chain case also differ in the available spin correlation information. For the single-chain case, the relevant quantity is the angular distribution of the decay products. Here, the decaying particle can be the first particle or any of the intermediate particles in the decay chain. In order to have a non-uniform angular distribution for the decay products, the decaying particle needs be polarized. In addition, if the decaying particle has spin 1/2, the coupling responsible for the decay needs be chiral. Of course, this information is also available in the double-chain case. On the other hand, if both decay chains are reconstructed, we obtain extra information unavailable in the single-chain case, namely, the spin correlations between the two decay chains.  

In this paper, we focus on the cases where the system is exactly-solvable or over-constrained\footnote{For an under-constrained system, although the missing momenta cannot be fully reconstructed, correlations between spin and kinematics often exist, see, for example, Ref.~\cite{Barr:2005dz, maos, Horton:2010bg}.}. In particular,  we analyze in detail decay chains with three visible particles, in both the single-chain case and the double-chain case. We will discuss the corresponding event reconstruction methods and related issues. It is straightforward to generalize the methods to other event topologies.
 
\section{Single Chain Technique}
\label{sec:single}
\subsection{Angular distribution of decay products}
\begin{figure}
\begin{center}
 \includegraphics[width=0.5\textwidth]{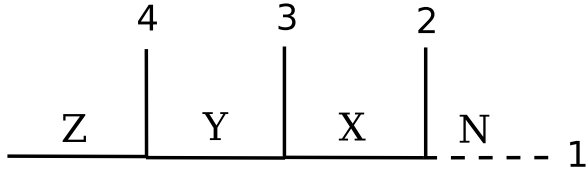}
\caption{\label{fig:single_chain} A decay chain with 3 visible SM particles. The final-state particles are labeled 1 through 4 with 1 denoting the missing particle and 2, 3, 4 denoting visible SM particles. The new particles are called $Z$, $Y$, $X$ and $N$ and assumed on shell.} 
\end{center}
\end{figure} 

A decay chain with 3 on-shell decays is shown in Fig.~\ref{fig:single_chain}. The particles $Z$ (not to be confused with the $Z$ boson, which we denote $Z_\mu$), $Y$, $X$ and $N$ are assumed to be on-shell with masses $m_Z$, $m_Y$, $m_X$ and $m_N$. Before describing the details of event reconstruction for this topology, we first discuss how to observe the spin correlation in the single-chain case once the missing particle's momentum is obtained, and compare it with the invariant mass method studied in the literature.

The basic idea for observing spin correlation from a particle decay is as follows: suppose the decaying particle is polarized and the coupling responsible for the decay is chiral, then the decay products will have a non-uniform angular distribution in the rest frame of the decaying particle. The two daughter particles' momenta are back to back in the rest frame of the mother particle and we use $\theta=\theta(m,d)$ to denote the angle between either of them and the polarization axis of the mother particle. Here $m$ denotes the mother particle and $d$ denotes the daughter particle. The probability density of the decay is a polynomial in $\cos\theta$ of order $2S$, where $S$ is the spin of the mother particle. Note that the coefficients of the polynomial depend on the spin density matrix of the decaying particle, the coupling responsible for the decay and also the axis one chooses to evaluate the angle $\theta$. In special cases, for example, when a fermion decays through a vector-like coupling, the coefficient(s) of the leading order term(s) could be vanishing or too small, giving a polynomial of order lower than $2S$. Therefore, using this method we can only set a lower bound on the decaying particle's spin. When the polarization axis is coincident with the direction of the mother particle's initial momentum, we say that the particle is polarized in the {\it helicity basis}, and denote the angle defined above by $\theta_{hel}(m,d)$. With full reconstruction of the kinematics of the event, $\theta_{hel}(m,d)$ can be simply obtained by boosting to the rest frame of the mother particle.  

\begin{figure}
\begin{center}
 \includegraphics[width=0.4\textwidth]{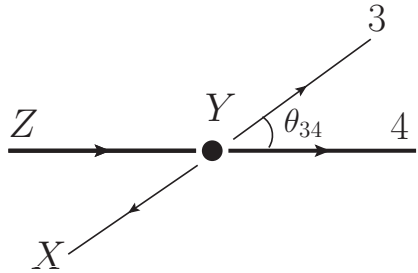}
\caption{\label{fig:invmass} The decay $Z\rightarrow Y\rightarrow X$ viewed in the $Y$ rest frame. }
\end{center}
\end{figure}

One may get a polarized particle if itself comes from the decay of another particle through a chiral vertex.
For example, we consider the particle $Y$ in Fig.~\ref{fig:single_chain}, which comes from a two-body decay of the particle $Z$. It then decays to two particles, $X$ and $3$. In the rest frame of $Z$, particle $Y$ and particle $4$ move in opposite directions and $Y$ is polarized along that direction. Now we can boost the system to the rest frame of $Y$ (Fig.~\ref{fig:invmass}). In this frame one can see that the angle between particles $3$ and $4$ is simply $\theta_{34}=\pi -\theta_{hel}(Y,3)$ since the direction of the particle $4$ is unchanged under the boost.  
The combined invariant mass of visible particles $4$ and $3$ can be easily calculated in this frame and it is related to the angle $\theta_{hel}(Y,3)$ by
\begin{equation}
m_{34}^2=(p_3+p_4)^2 =\frac12(m_{34}^{\mbox{max}})^2(1-\cos\theta_{34})=\frac12(m_{34}^{\mbox{max}})^2(1+\cos\theta_{hel}(Y,3)), 
\end{equation} 
where 
\begin{equation}
(m_{34}^{\mbox{max}})^2 =\frac{(m_Z^2-m_Y^2)(m_Y^2-m_X^2)}{m_Y^2}.
\end{equation}
This fact has been used for spin measurements in Ref.~\cite{Barr:2004ze, Smillie:2005ar, Datta:2005zs, Alves:2006df, Athanasiou:2006ef, Wang:2006hk, Smillie:2006cd, Kilic:2007zk, Csaki:2007xm,Wang:2008sw, Burns:2008cp, Gedalia:2009ym, Ehrenfeld:2009rt, Kong:2010mh}. The advantage of the invariant mass method is that the distribution can be obtained {\it without} complete event reconstruction. However, it is also clear that it requires the particle to come from a heavier particle decay, hence it can only be applied to the spin determinations of the intermediate particles directly in a decay chain\footnote{Of course once the spins of the intermediate particles are known, we can determine whether the first
(last) particle is a boson or fermion.}. While the last particle never decays and it is hard to obtain its spin information directly, the spin of the first particle may be determined directly if its polarization already exists at the production level. There is no experimentally measurable Lorentz-invariant quantity related to the angle in which we are interested in this case. On the other hand, by reconstructing the missing particle's momentum, we can directly examine the angle $\theta_{hel}(m,d)$ of the first particle decay in a decay chain and extract its spin information, which we describe in the following subsections.

\subsection{Momentum reconstruction}
We now describe the momentum reconstruction for a decay chain as shown in Fig.~\ref{fig:single_chain}. The mass shell constraints give the following equations:
\begin{eqnarray}
p_1^2 & = & m_N^2,\nonumber \\
(p_1+p_2)^2 & = & m_X^2,\nonumber\\
(p_1+p_2+p_3)^2 & = & m_Y^2,\nonumber\\
(p_1+p_2+p_3+p_4)^2 & = & m_Z^2,\label{eq:one_chain} 
\end{eqnarray}
where $p_1$ is the four-momentum of the invisible particle, $p_2, p_3, p_4$ are the four-momenta of the visible SM particles, $m_N, m_X, m_Y, m_Z$ are the measured masses. It is easy to see by taking the differences that these equations can be simplified to 3 linear equations plus a quadratic equation for the invisible momentum. Therefore, the system of these equations always admits two solutions, with the number of real solutions being 0 or 2. The solutions can become complex if we use wrong combinations of the visible particles or the experimental smearing is too large. Such ``bad'' events or combinations should be eliminated by requiring the solutions to be real.  However, in the classes of models under consideration, there is always another decay chain beside the one we wish to study, which can also contain similar final state particles. Assigning a final state particle to the ``wrong" decay chain sometimes can also yield real solutions.  In practice, separating out such contaminations can be very challenging. In our study, we choose to accept all real solutions and add them with 
{\em equal weight}. As shown in our case study below, making such a choice does not prevent us from extracting spin information. A more careful treatment of such combinatorial contamination should be able to further enhance the spin differentiation power.

\begin{figure}
\begin{center}
\begin{tabular}{c}
\includegraphics[width=0.5\textwidth]{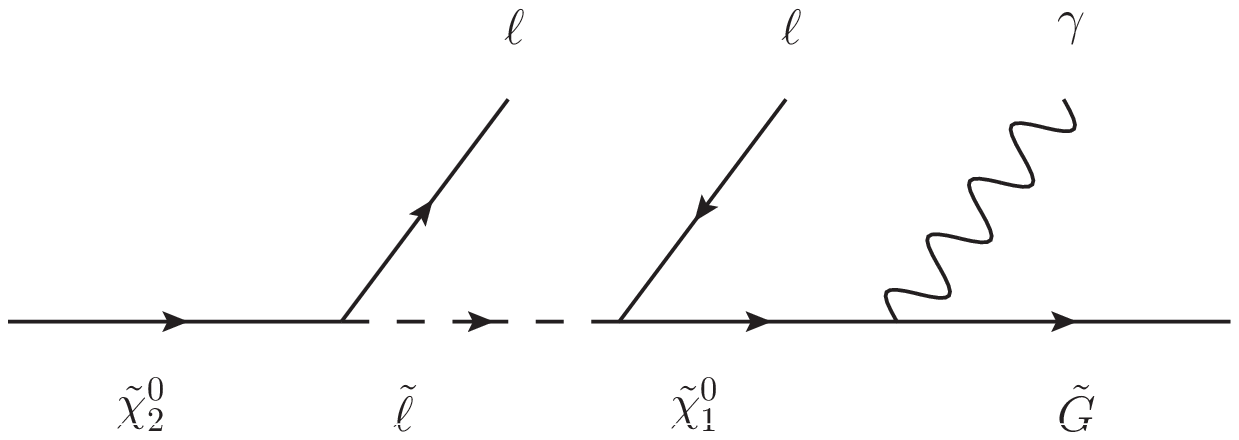}
\\(a) SUSY
\\
\includegraphics[width=0.5\textwidth]{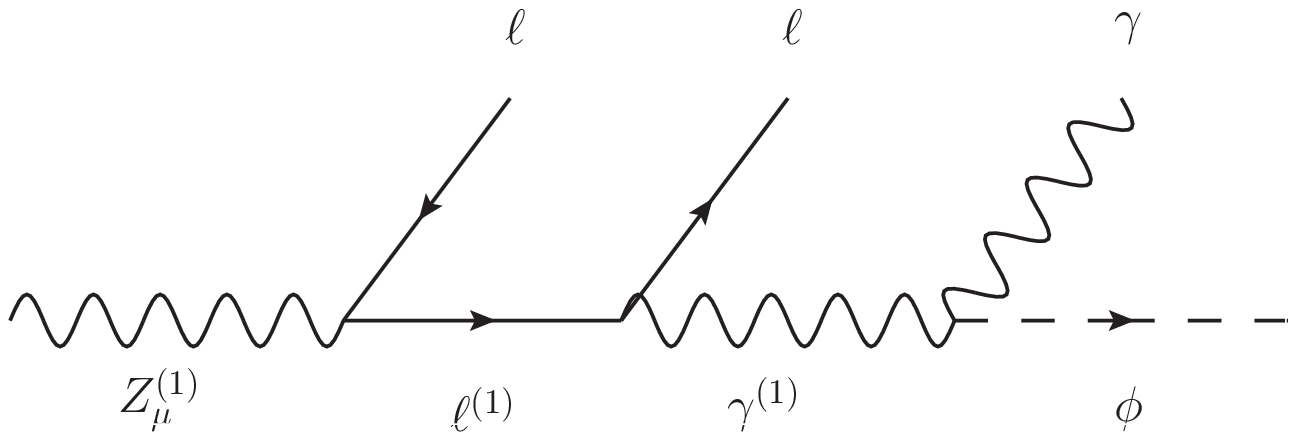}
\\(b) UED
\end{tabular}
\caption{Decay chains with photons in GMSB and UED.\label{fig:susy_ued}}
\end{center}
\end{figure}
\subsection{Application to decay chains with photons}
For an illustration, we apply the above method on a decay chain in a gauge mediated SUSY breaking (GMSB) model and its UED counterpart. These models are chosen simply because they give rise to the event topologies which are suitable for the spin studies with our method. GMSB is characterized by a gravitino LSP (lightest supersymmetric particle). We assume that the gravitino has a mass $\sim$eV which is essentially massless compared with the detector resolution. The next lightest supersymmetric particle (NLSP) is assumed to be a Bino-like neutralino which decays promptly to a photon and the gravitino. We are interested in the following decay chain: $\cntwo\rightarrow \ell_R\slep_R\rightarrow \ell_R\ell_R\cnone\rightarrow \ell_R\ell_R\gamma\gr$ (Fig.~\ref{fig:susy_ued} (a)), where $\cntwo$, $\slep$, $\cnone$ and $\gr$ are respectively the second neutralino, the right-handed slepton, the lightest neutralino and the gravitino, corresponding to the particles $Z$, $Y$, $X$ and $N$ in Fig.~\ref{fig:single_chain}. We set the mass of the gravitino to be zero and choose the other particles' masses and interactions according to the Minimal Supersymmetric Standard Model (MSSM) point SPS1a\footnote{SPS1a is actually a model point of the minimal supergravity (mSUGRA) instead of GMSB. Note also that Tevatron results have put stringent constraints on events with photons plus missing transverse momentum \cite{D0, CDF}, essentially ruling out a spectrum with such low masses. We just use this spectrum for illustration because of its clean chiral structure. It is straightforward to apply the same method on other model points.} \cite{snowmass}. Thus the masses of the particles $Z$, $Y$, $X$ and $N$ are 181, 143 , 97 and 0 GeV. As denoted by the subscripts, $\cntwo$ only decays to right-handed leptons. Therefore, if $\cntwo$ is polarized, we will be able to determine its spin. 

For comparison, we consider a similar decay chain in UED, $Z_\mu^{(1)}\rightarrow \ell_L\ell_L^{(1)}\rightarrow \ell_L\ell_L \gamma^{(1)}\rightarrow \ell_L\ell_L \gamma \phi$, where $Z^{(1)}_\mu$, $\ell_L^{(1)}$, $\gamma^{(1)}$ denote the first KK modes of the gauge bosons and leptons, and $\phi$ is a scalar field and the lightest KK-odd particle (LKP). Note that $\phi$ is absent in the minimal 5-dimensional (5D) UED model, but exists in 6D UED models \cite{6dued} as a scalar KK partner of the hypercharge gauge boson (dubbed $B_{H}^{(1,0)}$ in Ref.~\cite{6dued}). A scalar LKP also exists in an extension of the 5D UED model with an additional gauged Peccei-Quinn (PQ) $U(1)$ symmetry \cite{pqued}, as the zero mode scalar partner of the PQ gauge boson (denoted $B_5$). If the NLKP (next lightest KK-odd particle) is the KK-photon, it decays to a photon (or $Z$-boson) and $B_5$. Since that $B_5$ can be very light ($\lesssim$ GeV) while $B_{H}^{(1,0)}$ in the 6D UED model has to be massive ($\gtrsim$ 100GeV), the PQ-UED model can mimic more closely the signature of GMSB than the  6D UED model. For our purpose, the model subtleties are unimportant and we only need to fix the spins according to the model and specify the masses and couplings of the particles. We will choose the masses to be the same as the SUSY case. For simplicity, we assume $Z_\mu^{(1)}$ is purely $W^{3{(1)}}_\mu$ so that its couplings to fermions are purely left-handed:
\begin{equation}
\mathcal{L}\supset\bar q\gamma^\mu\frac{1-\gamma^5}{2} q^{(1)} Z^{(1)}_\mu + \text{h.c.} \label{eq:chiral_coupling}
\end{equation}
The KK-photon decays to $\phi$ through a vertex in the form
\begin{equation}
\mathcal{L}\supset c\epsilon^{\mu\nu\rho\sigma}\phi F_{\mu\nu}^{(1)} F_{\rho\sigma}^{(0)},\label{eq:kkphoton-phi}
\end{equation}
where $c$ is a coupling constant.

The process under consideration is the neutralino/chargino ($\cntwo$/$\cpmone$) pair production in GMSB and the UED counterpart KK-$Z_\mu$/KK-$W_\mu$ ($Z_\mu^{(1)}$/$W^{\pm (1)}_\mu$) pair production. Thus $\cntwo$/$Z_\mu^{(1)}$ is the first particle in the decay chain. In the lab frame, $\cntwo$ is more left-handed. This can be seen as follows: we consider the process $u\bar d\rightarrow \cntwo\cpone$, the other process $\bar u d\rightarrow \cntwo\cmone$ is similar. The process is dominated by the s-channel $W^+_\mu$-exchange diagram so that the initial $u$($\bar d$) is left-handed (right-handed). Therefore, in the center-of-mass frame, $\cntwo$ is more left-handed (right-handed) in the forward (backward) direction with respect to the $u$ quark due to angular momentum conservation.  In the center-of-mass frame, $\cntwo$ is equally left-handed and right-handed\footnote{This is not exactly true---$\cntwo$ is slightly polarized in the center-of-mass frame due to its small Higgsino component, see Ref.~\cite{Kitano:2008sa}}. However, the system tends to have a large boost along the $u$ quark direction, changing some of the right-handed $\cntwo$ to left-handed. Therefore we have more left-handed $\cntwo$ than right-handed in the lab frame. Similarly, in the UED case, KK-$Z_\mu$ is negatively polarized in the lab frame. 

Wrong combinations involving particles in the other decay chain can also yield real solutions and contaminate the distributions.
For SPS1a, 95\% of $\cpmone$ decays to a stau ($\stau$) and a neutrino. Therefore, we assume the chargino decays according to $\cpmone\rightarrow\stau\nu_\tau\rightarrow\tau\nu_\tau\cnone\rightarrow\tau\nu_\tau\gamma\gr$. Accordingly, we let KK-$W_\mu$ decay through a KK-$\tau$ to the KK-photon, which then decays to the scalar $\phi$ and a photon. Therefore, each event contains two photons, two opposite-sign same-flavor leptons and a (hadronic) $\tau$, which amount to a 4-fold ambiguity for assigning the positions of the two photons and the two leptons.  We will give a more detailed assessment of this combinatorial contamination in Sec.~\ref{subsec:combo}.

The events are generated with Herwig++ 2.4.2 \cite{Herwig++} at the parton level\footnote{Only the minimal UED is available in the official Herwig++ code. We have adjusted the code to allow a generic particle spectrum. We have also added the scalar field $\phi$ and let the KK-photon  decay through the coupling Eq.~(\ref{eq:kkphoton-phi}). The decay $\cnone\rightarrow \gamma\gr$ is performed according to phase space since the decay products have a uniform angular distribution.} for 14 TeV $pp$ collision. For simplicity, we have turned off initial/final state radiations. The final state radiation is small since the final state particles are either leptons or photons. The effect of the initial state radiation is to give the whole $\cntwo$/$\cpmone$ or KK-$Z_\mu$/KK-$W_\mu$ system a boost, which do not change qualitatively any of the results presented below.    

\begin{figure}[h!]
\begin{center}
\begin{tabular}{cc}
\includegraphics[width=0.5\textwidth]{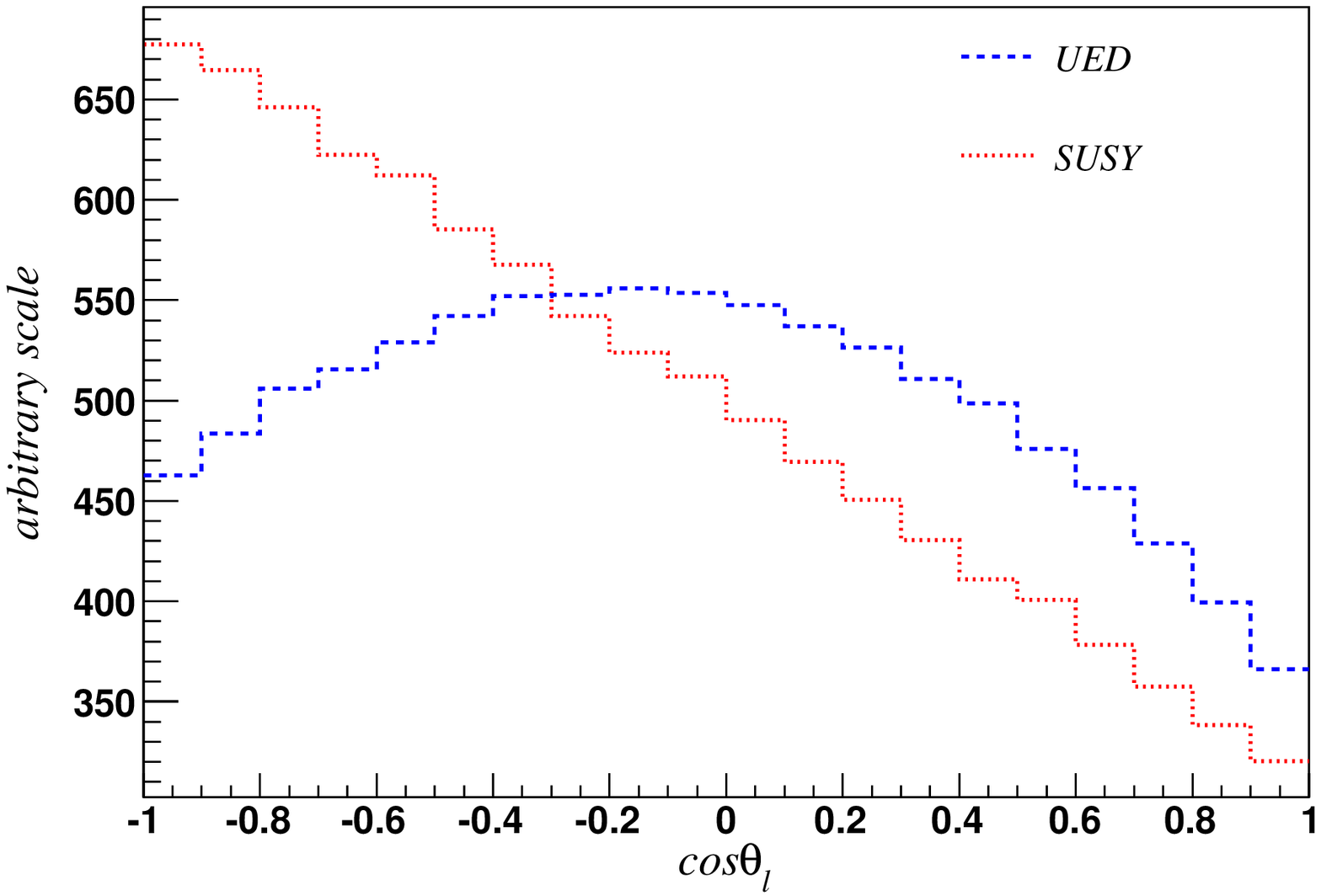}
&
\includegraphics[width=0.5\textwidth]{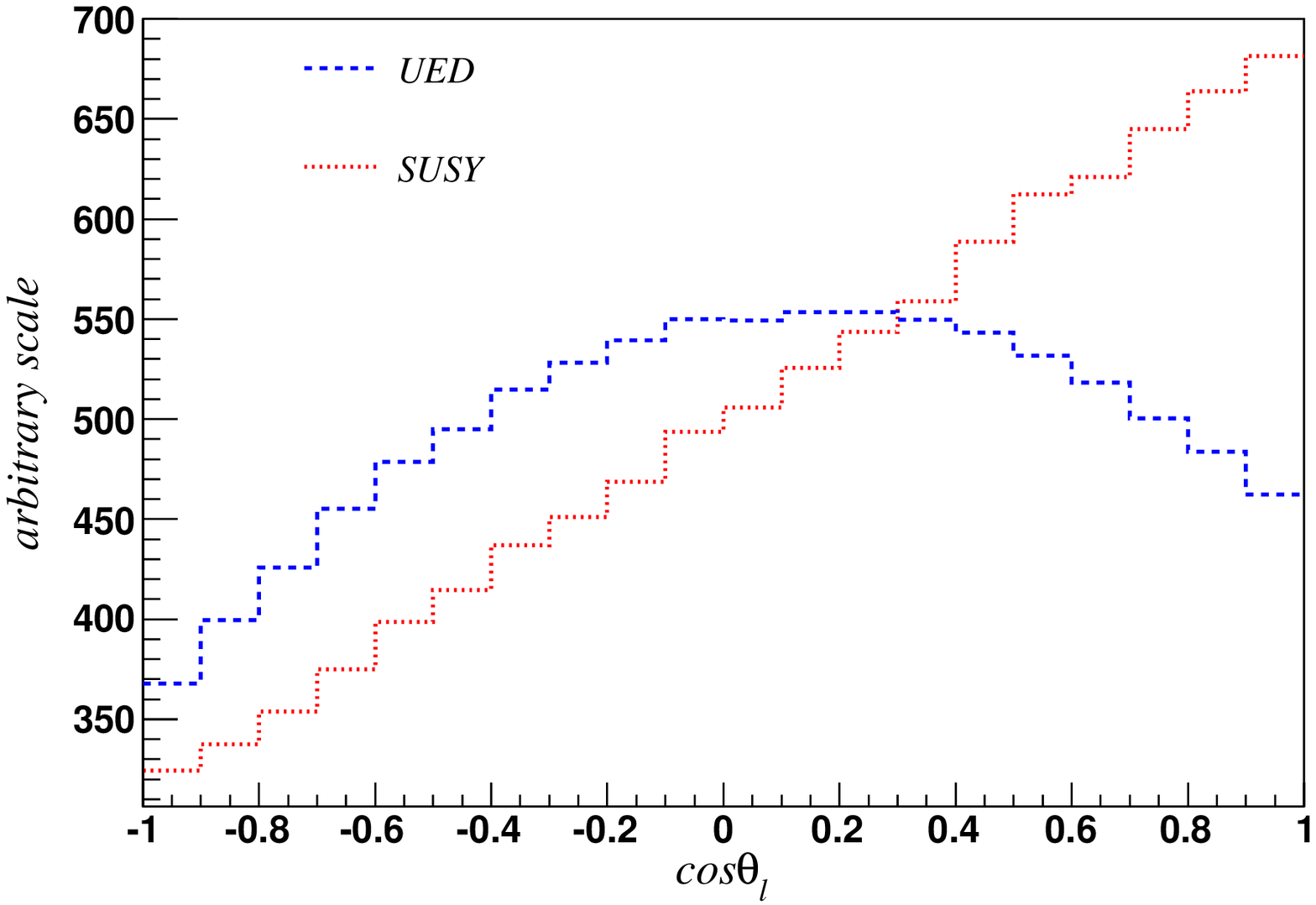}\\
(a)&(b)
\end{tabular}
\caption{Angle between the near lepton and $\cntwo$/$Z_\mu^{(1)}$ from Monte Carlo truth. Left: $\cos\theta_{\ell^+}$ for UED and $\cos\theta_{\ell^{-}}$ for SUSY; right: $\cos\theta_{\ell^-}$ for UED and $\cos\theta_{\ell^{+}}$ for SUSY. The number of events is normalized to 10k in 20 bins for all histograms.\label{fig:single_mc}}
\end{center}
\end{figure}

We are interested in the spin of the first particle in the decay chain, {\it i.e.}, $\cntwo$/$Z_\mu^{(1)}$. Therefore, we examine the angle $\theta_{\ell} \equiv \theta_{hel}(Z,\ell)$ as discussed above, where $\ell$ is the ``near'' lepton, namely, the lepton directly from the $\cntwo$/$Z_\mu^{(1)}$ two-body decay. In Fig.~\ref{fig:single_mc}, we show the distributions for $\cos\theta_{\ell}$ from the Monte Carlo.  For completeness, we draw the distributions separately for positive and negative near leptons . Because the coupling in the SUSY case is right-handed while in the UED case is left-handed,  the distribution of $\theta_{\ell^+}$ ($\theta_{\ell^-})$ in SUSY should be compared to that of $\theta_{\ell^-}$ ($\theta_{\ell^+}$) in UED, which we put in the same figure. It is clear that the distribution is linear for SUSY and quadratic for UED, corresponding to spin-1/2 and spin-1 particles. We have normalized the number of events for each distribution to 10k (20 bins), although we have used more events to produce the smooth distributions.  Note that it is unnecessary to make this distinction based on lepton charge, if we limit our goal to differentiate SUSY from UED, since we only need to distinguish linear vs quadratic behavior in this case. On the other hand, the slope does carry the information of chirality of the coupling in this case.

\begin{figure}
\begin{center}
\begin{tabular}{cc}
\includegraphics[width=0.5\textwidth]{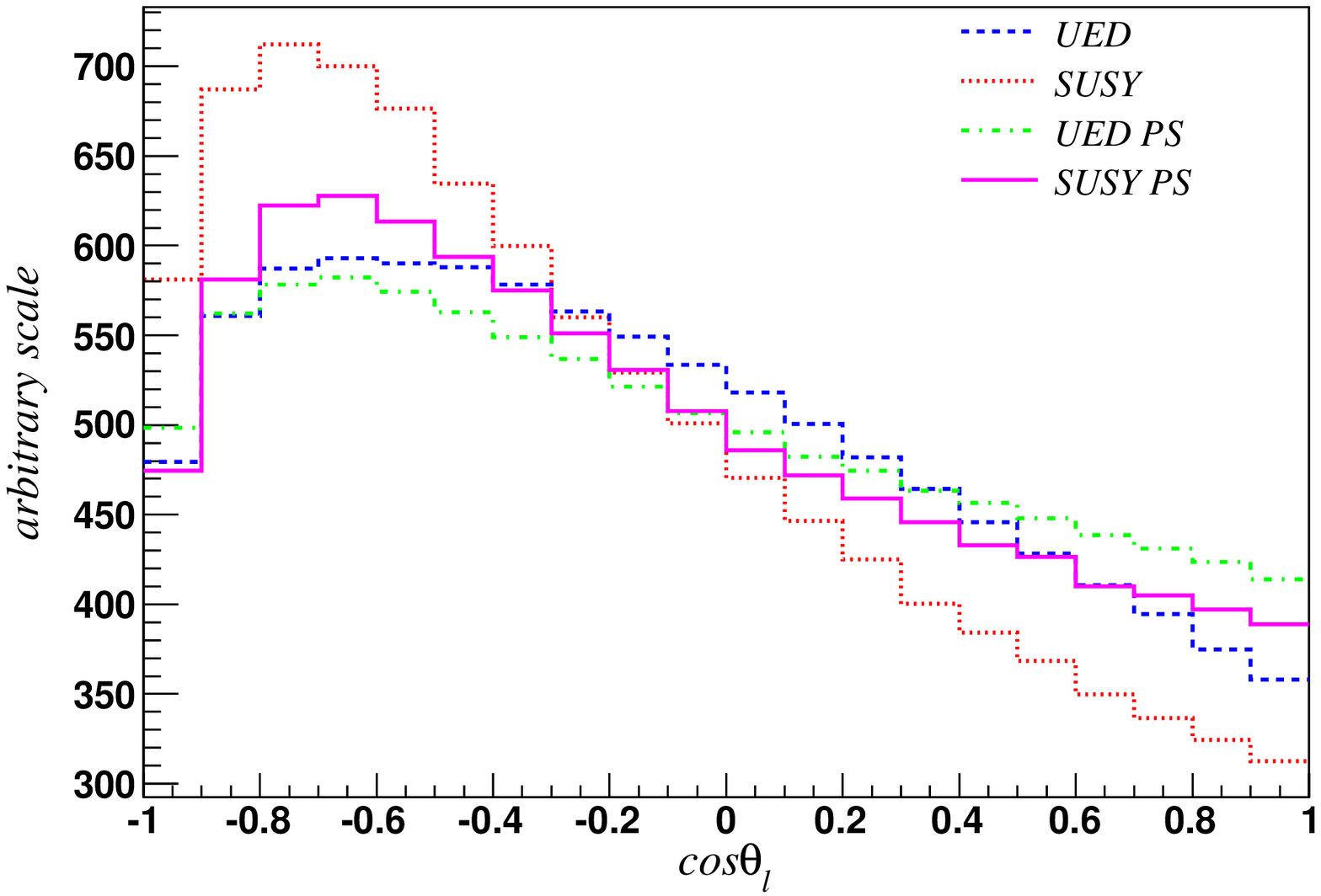}
&
\includegraphics[width=0.5\textwidth]{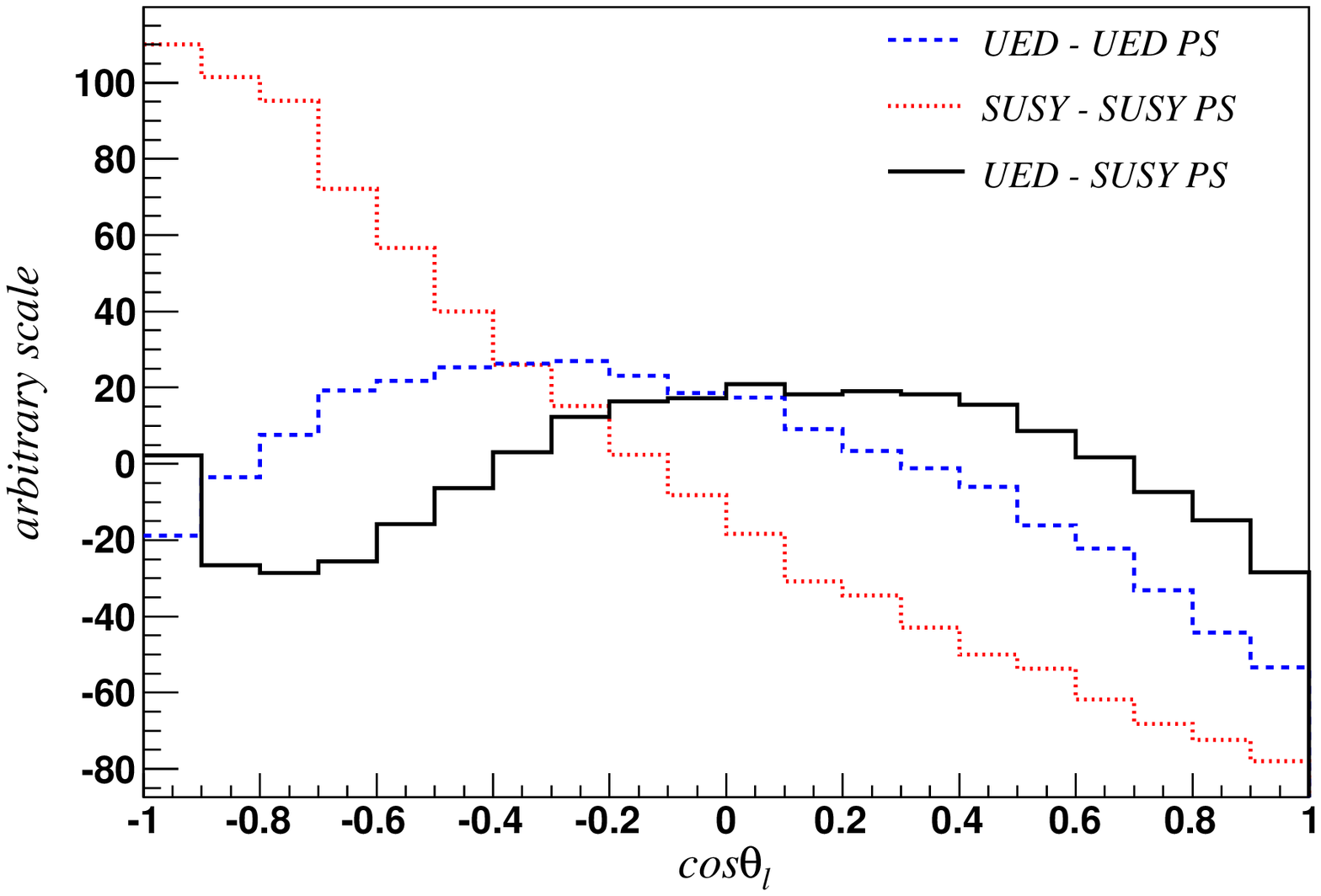}\\
(a)&(b)
\end{tabular}
\caption{Reconstructed $\cos\theta_\ell$. Exact momenta without experimental smearing are used. All combinations included. Only showing   $\cos\theta_{\ell^+}$ for UED and $\cos\theta_{\ell^{-}}$ for SUSY. The UED (SUSY) PS distribution is obtained using the UED (SUSY) $2\rightarrow2$ differential cross-section with all decays performed according to phase space. Left: before subtracting PS distributions; right: after subtracting PS distributions. We normalize the number of solutions for all histograms on the left panel to 10k, and do the subtractions to obtain the distributions on the right panel without further normalization.  \label{fig:single_rec}}
\end{center}
\end{figure}
We first apply the event reconstruction method on events without any experimental cuts or smearing. 
The distributions including all solutions with equal weight are shown in Fig.~\ref{fig:single_rec} (a). For comparison, we have also performed the reconstruction for events with the same mass spectrum and $2\rightarrow2$ differential cross-section (for UED and SUSY respectively), but with all particles decayed according to phase space (PS). Comparing Fig.~\ref{fig:single_rec}(a) and Fig.~\ref{fig:single_mc}, one can see that the distributions have been distorted from the Monte Carlo truth by wrong combinations and wrong solutions, and the theoretical linear and quadratic functions of Fig.~\ref{fig:single_mc} are lost. Nevertheless, the distributions of UED, SUSY and PS are clearly distinguishable.  We can also retrieve (some of) the theoretical behavior by subtracting the UED and SUSY distributions from the corresponding PS ones, which are shown in Fig.~\ref{fig:single_rec} (b). The subtracted distributions are much closer to the original ones, although the contamination cannot be completely removed.  
Note that SUSY and UED give rise to different PS distributions, which can be attributed to the difference in the differential production cross section, as discussed in more detail later in Sec.~\ref{subsec:combo}. 
A further potential obstacle is that, in practice, we do not know which PS distribution to compare to. However, as shown in Fig.~\ref{fig:single_rec}(b), the distributions are still distinguishable even if we made the wrong subtraction.  

\begin{figure}
\begin{center}
\begin{tabular}{cc}
\includegraphics[width=0.5\textwidth]{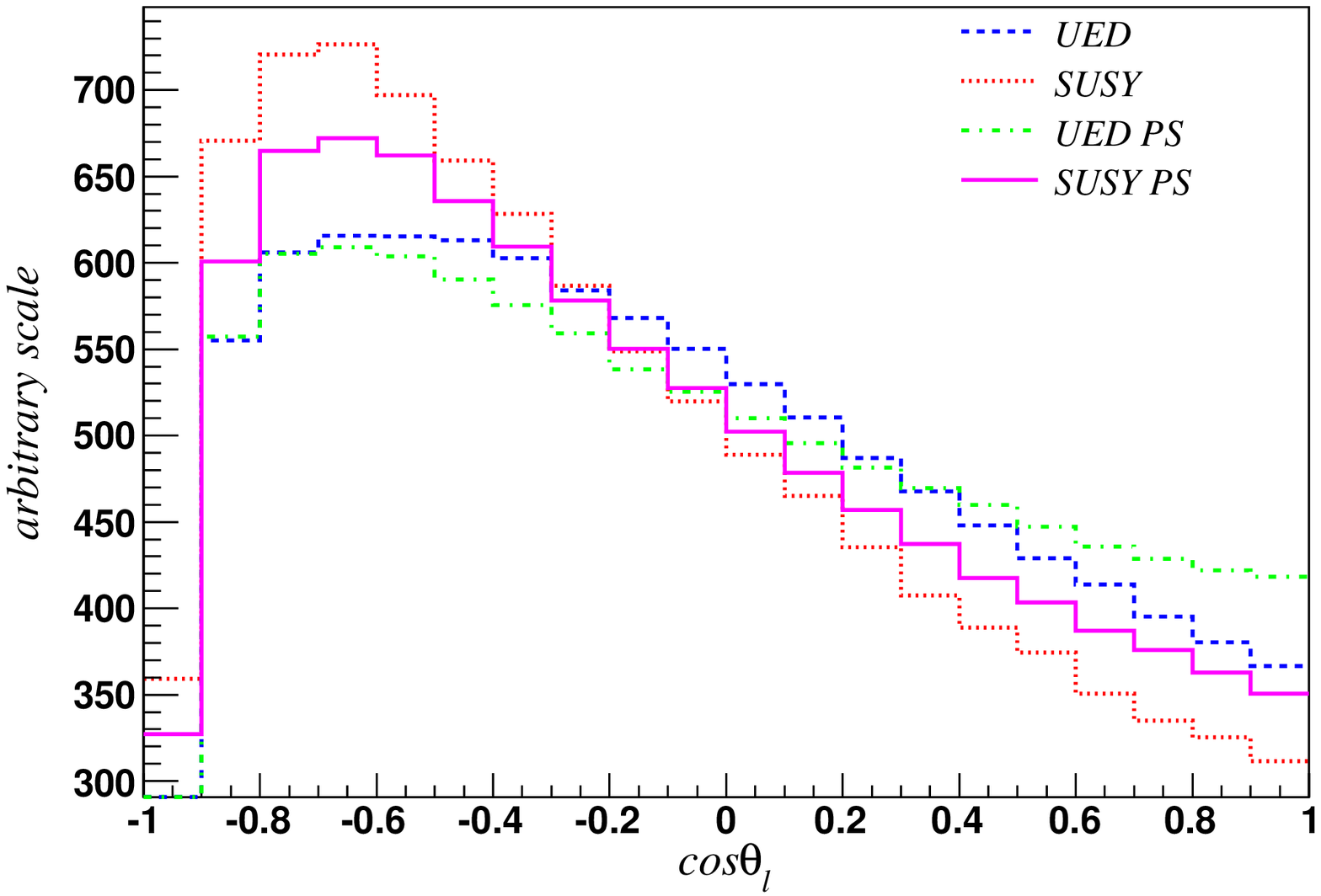}
&
\includegraphics[width=0.5\textwidth]{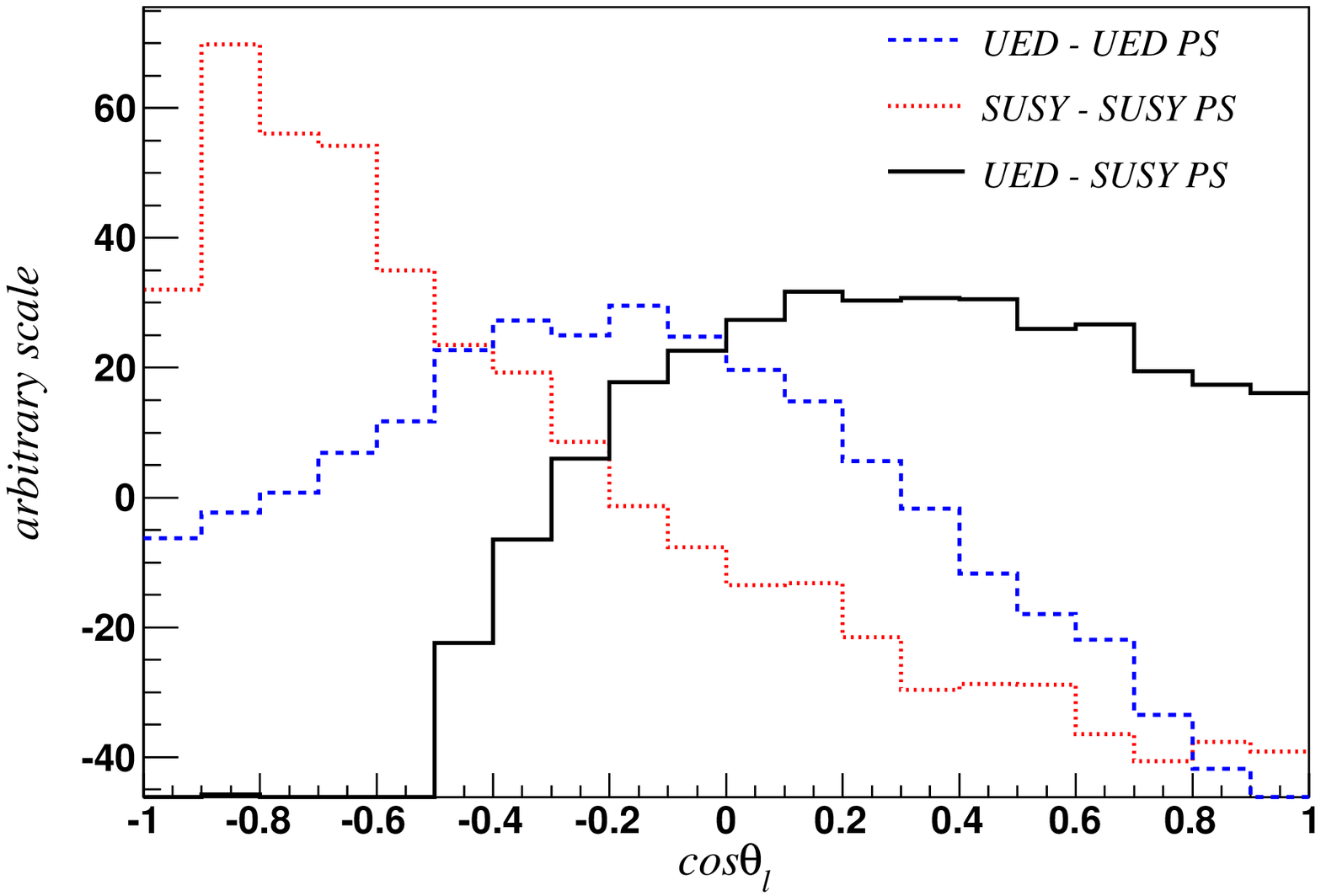}\\
(a)&(b)
\end{tabular}
\caption{Reconstructed $\cos\theta_\ell$. Experimental smearing applied. All combinations included. Only showing $\cos\theta_{\ell^+}$ for UED and $\cos\theta_{\ell^{-}}$ for SUSY. Left: before subtracting PS distribution; right: after subtracting PS distribution. We normalize the number of solutions for all histograms on the left panel to 10k, and do the subtractions to obtain the distributions on the right panel without further normalization.\label{fig:single_smeared}}
\end{center}
\end{figure}

The actual distribution observed in a collider detector is also subject to modifications from experimental smearing, cuts, efficiency, {\it etc}. We simulate the detector response using a simplified approximation described in the Appendix, 
taking into account the detector coverage and momentum resolution\footnote{ Of course, our detector simulation is far from a complete one, which has to include effects such as trigger efficiency, mis-identification rate, isolation cuts, etc.}. The cuts on $p_T$ ($> 10$ GeV for both leptons and photons) and $|\eta|$ ($< 2.4$ for leptons and $< 3.0$ for photons) reduce the number of events to 82\% for UED and 63\% for SUSY. The UED efficiency is larger because the particles have higher $p_T$, as explained later in Sec.~\ref{subsec:combo}. Since the visible particles in our example are either leptons or photons, both of which have good resolutions in a collider detector, the experimentally smeared distributions (Fig.~\ref{fig:single_smeared}) are not significantly different from those using exact momenta.

\begin{figure}
\begin{center}
\begin{tabular}{cc}
\includegraphics[width=0.5\textwidth]{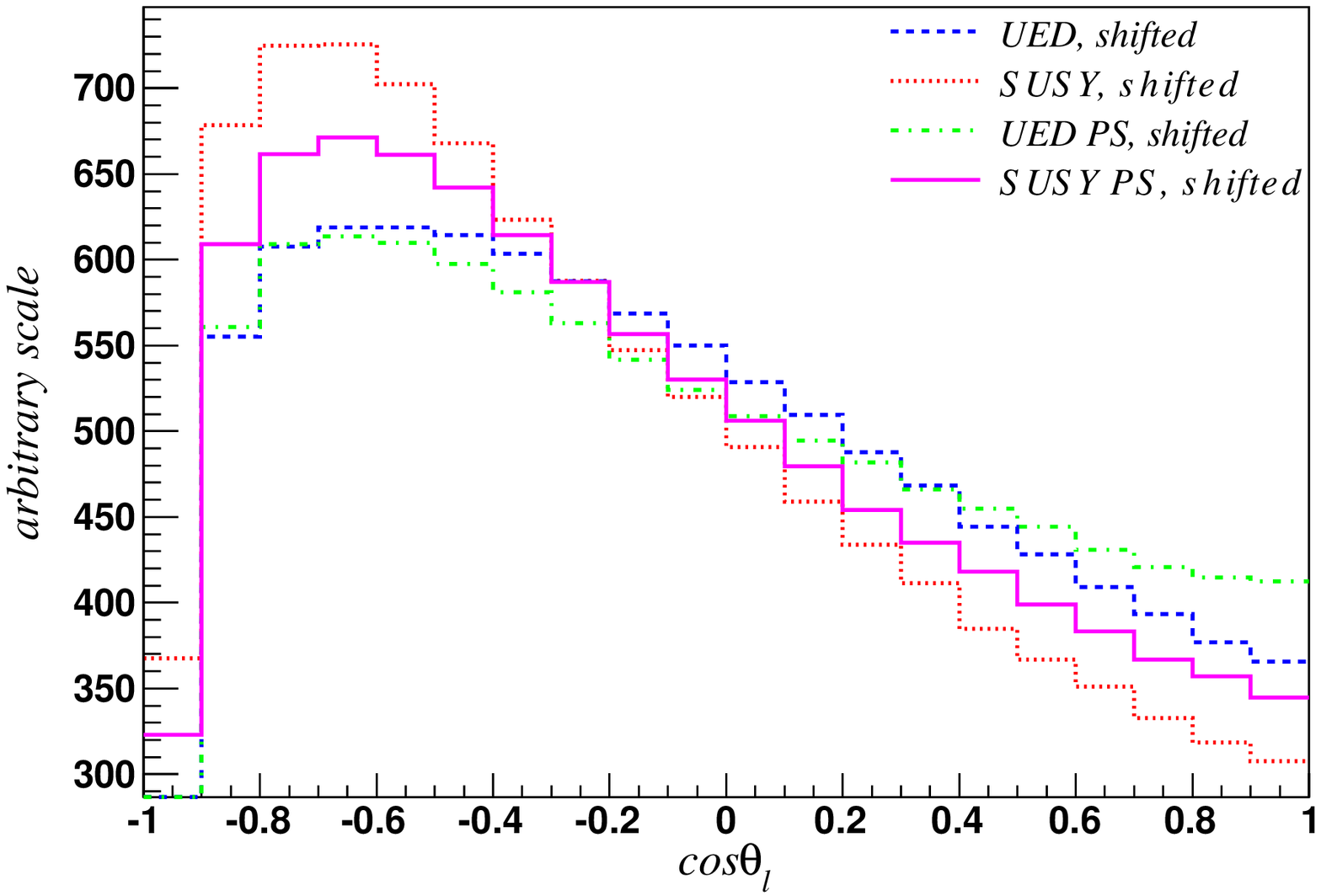}
&
\includegraphics[width=0.5\textwidth]{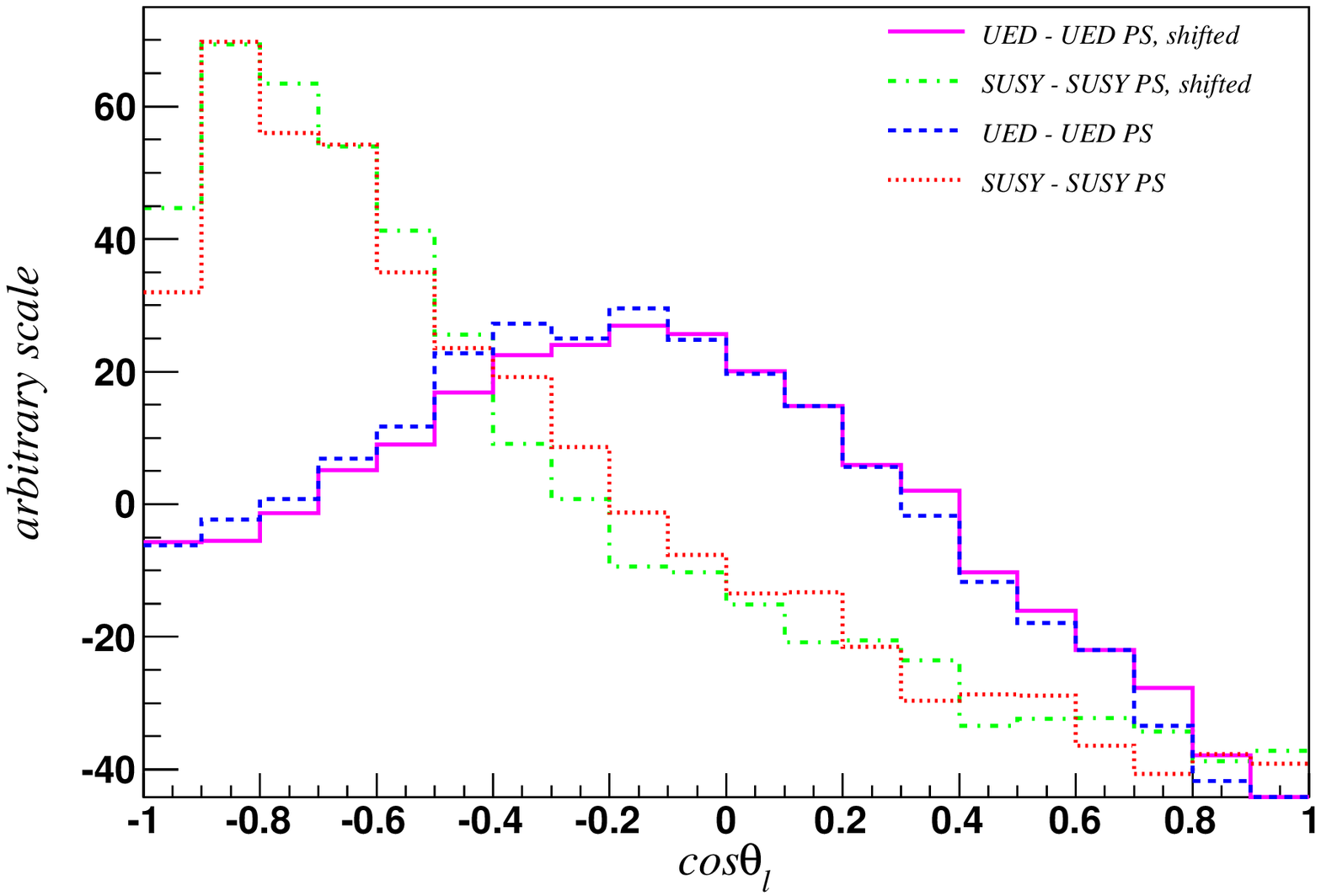}\\
(a)&(b)
\end{tabular}
\caption{\label{fig:single_shifted}The same plots as in Fig.~6, but with all input masses shifted by $+5$ GeV from the correct values. The original distributions in Fig.~6 (b) are superimposed in the right panel for comparison.}
\end{center}
\end{figure}

When producing Fig.~\ref{fig:single_smeared}, we have used the correct masses to obtain the solutions. In practice, the masses are measured with errors which could alter the distributions if they are significantly different from the correct values. However, for decay chains with multiple leptons and photons as we are considering, we expect good resolutions of mass measurements. To estimate the effect of mass measurement errors, we shift the input masses by $+5$ GeV and repeat the above procedure. The resulting distributions are given in Fig.~\ref{fig:single_shifted}, showing only tiny shifts from the distributions in Fig.~\ref{fig:single_smeared}. More importantly, it shows that such errors in mass measurements do not change the distinction between the SUSY and UED distributions.

\subsection{Combinatorics and more global information}
\label{subsec:combo}

\TABLE[ht!]{
\begin{tabular}{c|c|c|c|c}
\hline
     & correct & wrong photon & wrong lepton & wrong photon \& wrong lepton\\
\hline
UED  & 2 & 0.13 & 1.21 & 0.12\\
SUSY & 2 & 0.70 & 1.20 & 0.67\\
UED PS   & 2 & 0.13 & 1.21 & 0.12\\
SUSY PS  & 2 & 0.70 & 1.21 & 0.67\\
\hline
\end{tabular}
\caption{Average number of solutions per event for various combinations. Exact momenta without smearing are used. The correct combination always yields two solutions. ``Wrong photon'' means that we have used the photon from the wrong decay chain. ``Wrong lepton'' means that the two leptons in the decay chain are interchanged from their correct positions. \label{tab:solutions}}
}

\begin{figure}[h!]
\begin{center}
\begin{tabular}{cc}
 \includegraphics[width=0.5\textwidth]{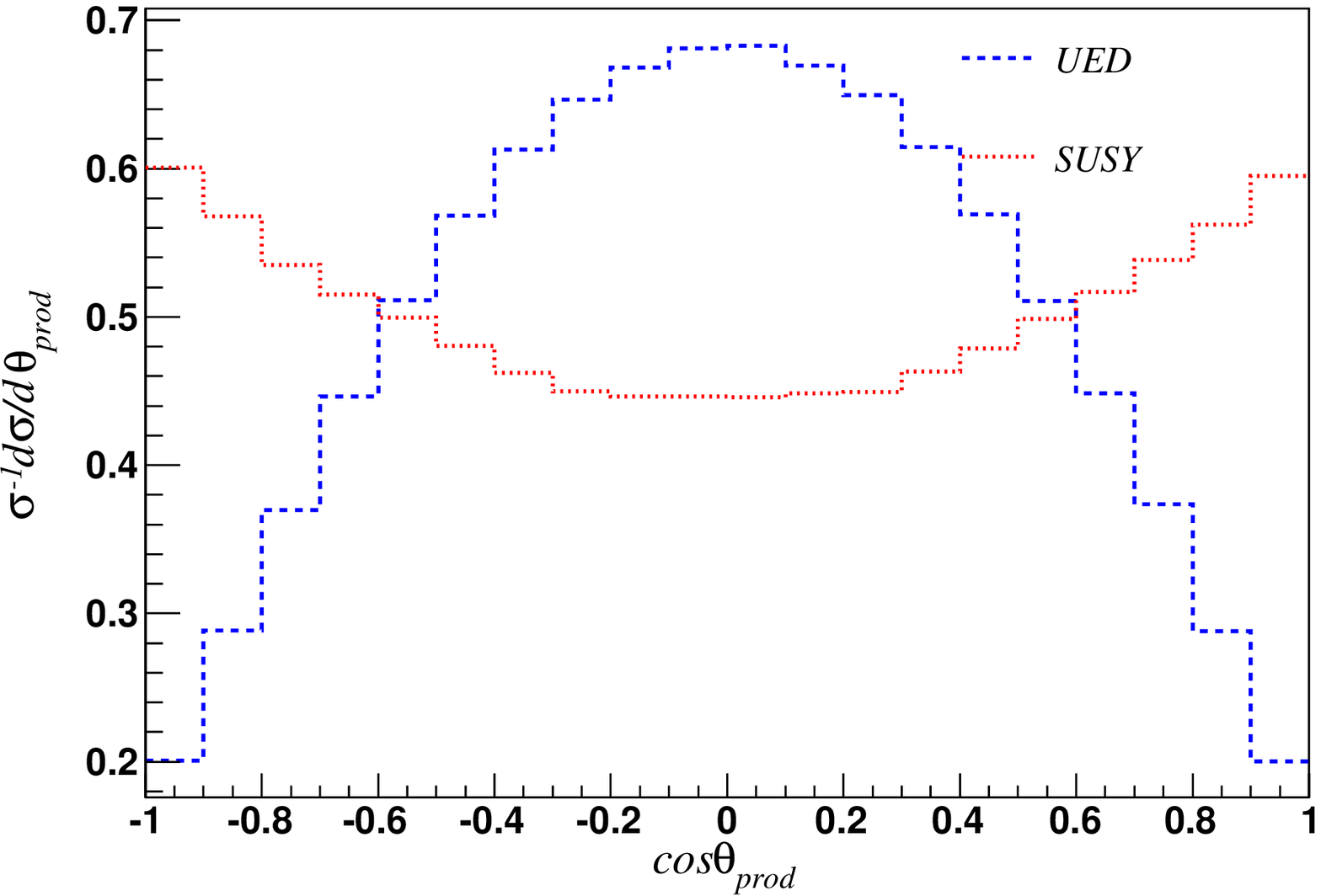} &
 \includegraphics[width=0.5\textwidth]{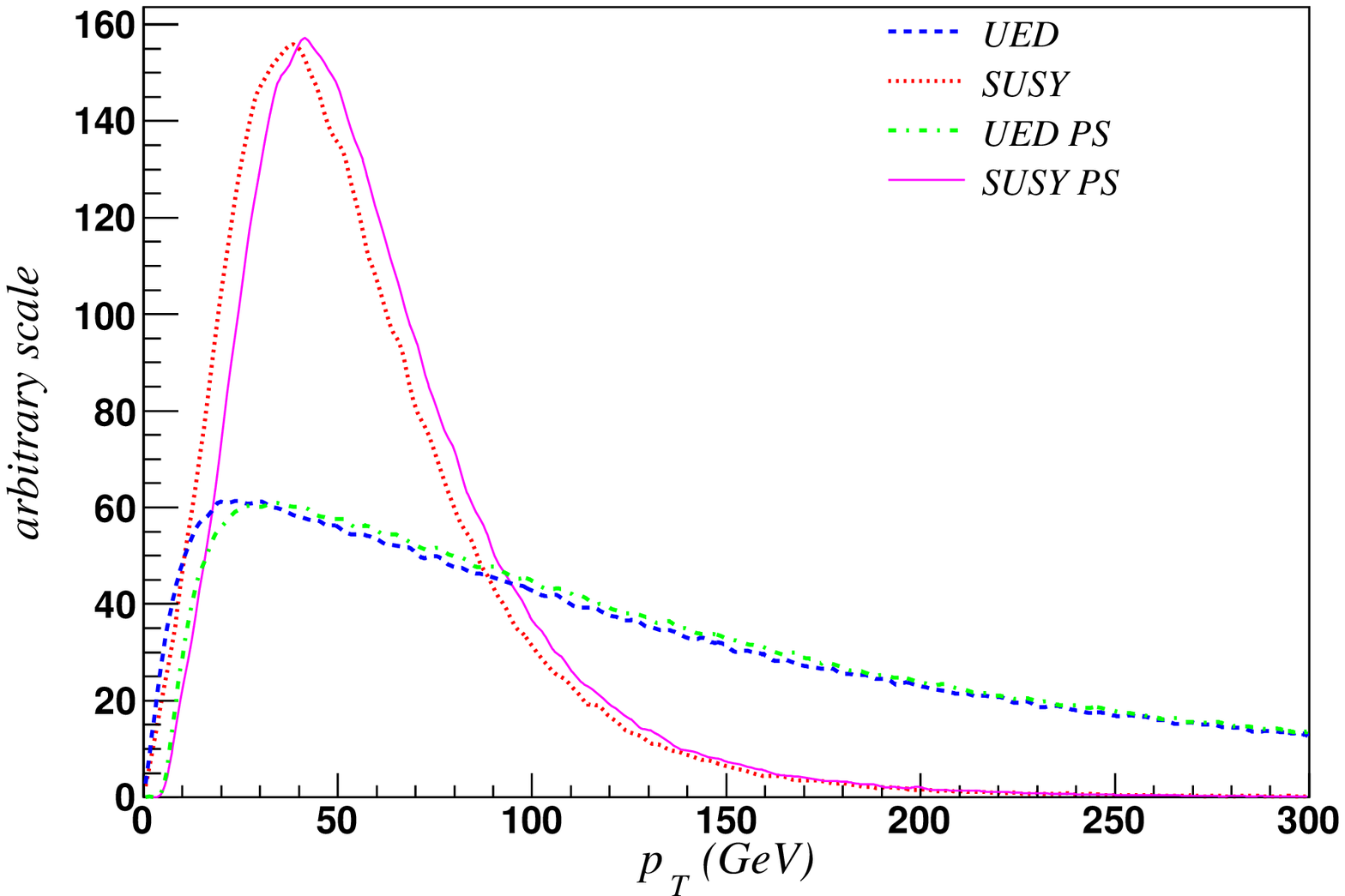} 
 \end{tabular}
\caption{Left: Production angle in the neutralino chargino center of mass frame. Right: $p_T$ distributions of the wrong photons. The UED PS (SUSY PS) distributions are shifted horizontally by 5 GeV to distinguish from the UED (SUSY) distribution. \label{fig:pt_wrongphoton}} 
\end{center}
\end{figure}

It is interesting to examine the number of real solutions decomposed according to different combinations, which is shown in Table~\ref{tab:solutions}. As mentioned before, we can eliminate some wrong combinations by requiring the solutions to be real. Nevertheless, there are still significant contributions. From Table~\ref{tab:solutions}, we see that the number of solutions from wrong lepton combinations is similar for all cases. On the other hand, the number of solutions from wrong photons is sensitive to the kinematics of the other decay chain, in this case, the $p_T$ of the wrong photon (Fig.~\ref{fig:pt_wrongphoton}). From the right panel of Fig.~\ref{fig:pt_wrongphoton}, we see that the $p_T$ distribution of the wrong photon is almost identical for SUSY and SUSY PS (the same is true for UED and UED PS), but very different between SUSY and UED. This is due to the distinction in the $2\rightarrow 2$ differential cross-section for SUSY and UED. As shown in the left panel of Fig.~\ref{fig:pt_wrongphoton}, the productions in UED events  tend to be more central, leading to a harder photon spectrum.
 
It may be possible to develop more sophisticated methods to reduce the wrong combinations using additional information of the other decay chain. To achieve that more precise knowledge of all possible decay chains, such as masses and couplings of all particles involved, are often necessary.  However, the strategy would be highly model-dependent in order to be more effective. Although it could be a useful step in practice, we will not pursue this further complication in our analysis. 
 
Finally, we note that the difference in production angle shown in the left panel of Fig.~\ref{fig:pt_wrongphoton} is directly correlated with the difference in spin of the particles. Hence, this production angle itself is a very good variable for spin measurement.  However, in this particular channel, this angle is not readily reconstructable due to the presence of a neutrino in the chargino decay chain.  Although less direct, as shown above, the $p_T$ distribution of the wrong photon is sensitive to the production angle,  and it can be used for spin measurement. In the next section, we will explore cases where we can reconstruct both decay chains, then the production angle provides a direct probe of the spin. 

\section{Double Chain Techniques}
\label{sec:double}

\begin{figure}[h!]
\begin{center}
 \includegraphics[width=0.4\textwidth]{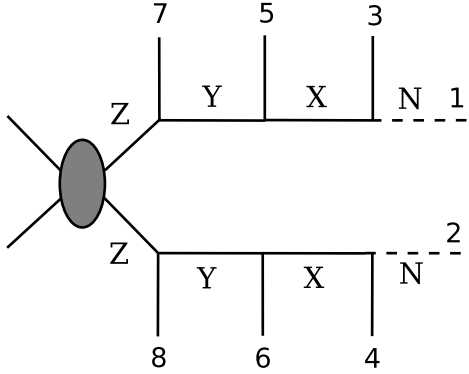}
\caption{\label{fig:double_chain} An event with two decay chains, each containing 3 visible SM particles. The final-state particles are labeled 1 through 8 with 1 and 2 denoting the two missing particles and 3-8 denoting visible SM particles. The new particles are called $Z$, $Y$, $X$ and $N$ and assumed on-shell.} 
\end{center}
\end{figure} 
In this section, we present techniques based on reconstruction of both of the decay chains. In principle, they are applicable to any event topology with enough constraints to solve the kinematics of both decay chains. As a demonstration, we focus on the case with two identical decay chains, as shown in Fig.~\ref{fig:double_chain}. Event reconstruction has been considered for this event topology in Refs.~\cite{mass33, Nojiri:2010dk}. The purpose there is to determine the masses of the particles in the decay chain, while the goal here is to find the best-fit momenta for the invisible particles assuming all masses are known (with uncertainties). The invisible particles' momenta are obtained as follows. First, we have 8 equations from the mass-shell constraints of the 8 on-shell particles in the two decay chains. In addition, if the only missing particles in the events are the two neutral particles at the end of the decay chains, we have two additional constraints,
\begin{equation}
p_1^x+p_2^x=\slashchar{p}^x,\ \ p_1^y+p_2^y=\slashchar{p}^y.
\end{equation} 
Therefore, we have 10 equations and 8 unknowns and the system is over-constrained. Using the uncertainties in mass measurement given in Refs.~\cite{mass33}, together with the experimental errors for the visible momenta, we perform a likelihood fit to find event by event the best fit momenta of the missing particles. We describe the fitting procedure in Appendix \ref{app:likelihood}.

Once the momenta of the missing particles are reconstructed, we can of course obtain the angular distributions of the decay products as in the single chain case. More interestingly, we also obtain information unavailable in the single chain case, which we illustrate by applying the method on sbottom/KK-bottom pair productions.

\subsection{Application to sbottom/KK-bottom pair productions}  
In this process, each event contains two sbottoms/KK-bottoms.  The sbottom is assumed to decay in the following decay chain:
\begin{equation}
\sbot\rightarrow b\cntwo\rightarrow b\ell\slep\rightarrow b\ell\ell\cnone, \label{eq:sbottom_decay}
\end{equation}   
and a similar decay chain occurs for the KK-bottom. Notice that we have enough constraints here to carry out a single chain analysis, using the method presented in Sec.~\ref{sec:single}. However, in this case, the single chain analysis will not reveal the spin of the sbottom or KK-bottom. The decay product of sbottom will give a flat distribution in $\theta_{hel}$. At the same time, the KK-bottoms are produced mostly through their couplings to gluon which is vector-like. Therefore, the KK-bottom is almost all unpolarized and the $\theta_{hel}$ distribution of the decay products is flat as well. Therefore, in this case, we can only get the spin information of the sbottom/KK-bottom from a double chain analysis. 

We use the SUSY particle spectrum of SPS1a, and we set the UED mass spectrum to be the same. The masses for particles $Z$, $Y$, $X$ and $N$, corresponding to $\tilde{b}$, $\tilde{\chi}_2^0$, $\tilde{\ell}$, and $\tilde{\chi}_1^0$ (or similar KK states), are then \{515, 180, 144, 97\} GeV. All UED couplings are assumed to be chiral according to Eq.~(\ref{eq:chiral_coupling}). The leading order cross-section is 0.36 pb for SUSY and 2.3 pb for UED. The actual event rate for both sbottoms/KK-bottoms to decay according to Eq.~(\ref{eq:sbottom_decay}) highly depends on the decay branching ratios. For UED, the KK-bottom has a branching ratio $\sim1/3$ to KK-$Z_\mu$ and KK-$Z_\mu$ has a branching ratio $\sim2/3$ to KK-$e$ or KK-$\mu$, therefore the effective cross-section is 0.11 pb.  For SPS1a, $\cntwo$ dominantly decays to the stau, and the branching ratio of $\cntwo\rightarrow\smu/\se$ is only 12\%, which makes the effective cross-section much smaller: $\sim 5.8$ fb. Of course, this suppression is not generic, and if necessary one can also consider the stau, though with less precision \cite{Godbole:2008it, Brooijmans:2010tn}. For a similar spectrum (with the 515 GeV sbottom replaced by a 565 GeV squark), it was shown in Ref.~\cite{mass33} that the masses can be determined with a few GeV uncertainties using 400 events, independent of the spins of the particles. For spin determination, the needed number of events is larger. Therefore, we simply use the errors given in Ref.~\cite{mass33} in our fit:
 \begin{equation}
\delta m_N = 4\gev, \delta m_X =4\gev, \delta m_Y = 4\gev, \delta m_Z = 6\gev.     
\end{equation}

We smear the visible particles' momenta according to Appendix \ref{app:likelihood}. About 74\% (72\%) events passed the $p_T$ and $\eta$ cuts for SUSY (UED).  We then apply the likelihood method described in Appendix \ref{app:likelihood} to reconstruct the momenta of the two missing particles. It allows us to obtain a minimum $\chi^2$ ($\chi^2_{\min}$) for each combination of the visible particles. For simplicity, we only keep the combination that gives the smallest $\chi^2_{\min}$, although sometimes more than one combinations yield good fits. A fit quality cut $\chi^2 < 10$ is applied on the events, which further reduces the number of events to 61\% (60\%) for SUSY (UED) with respect to the original number without cuts. Out of the final events after all cuts, about 46\% events (for both SUSY and UED) have the correct combination as checked with the event records from the Monte Carlo simulation. 

\begin{figure}[h!]
\begin{center}
 \includegraphics[width=0.8\textwidth]{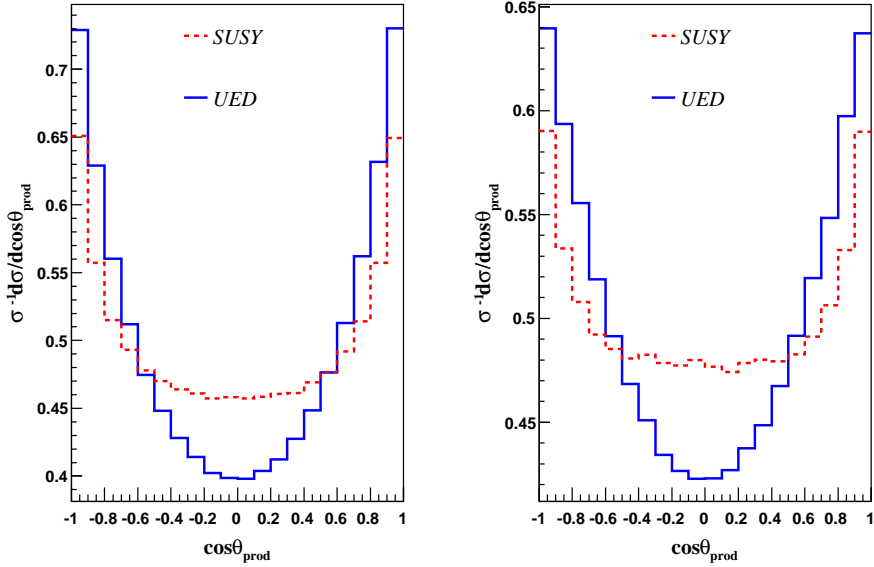}
\caption{\label{fig:production}The production angle in the center of mass frame of sbottoms/KK-bottoms. Left: Monte Carlo truth. Right: from reconstruction. } 
\end{center}
\end{figure}

 \begin{figure}[h!]
\begin{center}
 \includegraphics[width=0.8\textwidth]{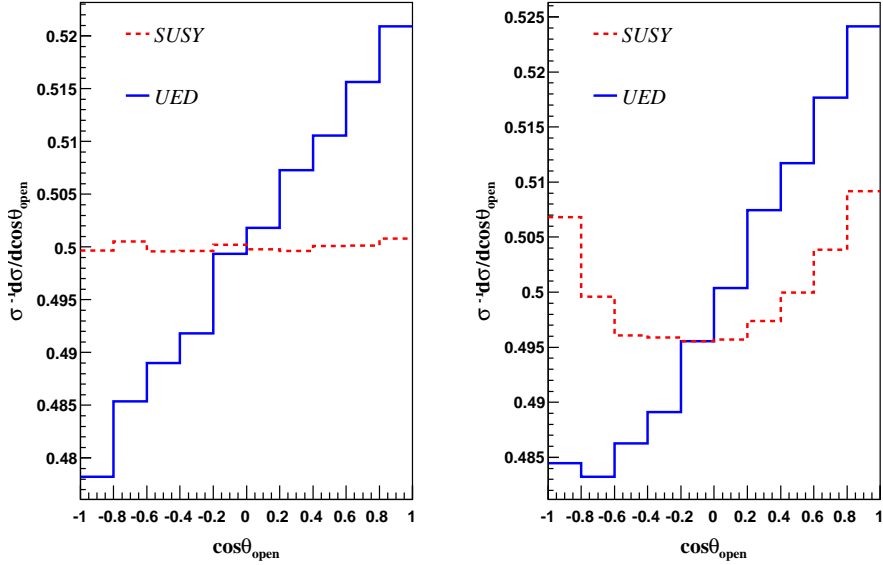}
\caption{The opening angle between the two b-jets, in the rest frames of the two KK-$b$'s respectively. Left: Monte Carlo truth; right: reconstructed.\label{fig:jj_open}} 
\end{center}
\end{figure}     

After obtaining the missing particles' momenta, we can calculate the momenta of the sbottoms/KK-bottoms and examine the production angle, {\it i.e.}, the angle between $\tilde b$/$b^{(1)}$ and the beam in the center of mass frame. The production angle distributions from both the Monte Carlo truth and the reconstruction are shown in Fig.~\ref{fig:production}. We see that although this distribution is useful to tell the two models apart, the shapes of the curves are not as distinct as those of $\tilde{\chi}^+ \tilde{\chi}_2^0$ and $W^{(1)+} Z^{(1)}$ productions studied in Section~\ref{sec:single}.   This is expected.  The $\tilde{\chi}^+ \tilde{\chi}_2^0$ and $W^{(1)+} Z^{(1)}$ production processes are dominated by $s$-channel $W^+$ processes from $u \bar{d}$ initial state. Therefore, it has a simple partial wave structure, and the spin of the final state particles in the $2 \to 2$ process determines the angular distribution. On the other hand, $\tilde{b} \tilde{b}^*$  and $b^{(1)} \bar{b}^{(1)}$ productions receive contributions from both $gg$ and $q \bar{q}$ initial states. In addition, $gg$ initiated production processes receive contributions from  $s$, $t$ and $u$ channels.  Therefore, the dependence on the spin of the final state particles in the $2 \to 2 $ process is weakened. 

Additional information can be obtained by studying spin correlation between the two decay chains. Of course, there is no correlation in the sbottom pair because they are scalar fields. On the other hand, we do expect correlations between the two KK-bottoms' helicities.  This is analogous to the $t\bar t$ spin correlation \cite{topspin} except that the mass is different. Therefore, we refer readers to Ref.~\cite{topspin} for the detailed discussion.  The correlation can be observed by examining the opening angle, $\theta_{\rm open}$ \cite{openangle}, between the two $b$-jets. $\theta_{\rm open}$ is defined as follows. We denote unit vector $\hat{p}_{b}$ as the direction of the bottom quark in the rest frame of $b^{(1)}$, and similarly $\hat{p}_{\bar{b}}$ as the direction of $\bar{b}$ in the rest frame of $\bar{b}^{(1)}$.  Obviously $\hat{p}_{b}$ and $\hat{p}_{\bar{b}}$ carry information of the polarizations of $b^{(1)}$ and $\bar{b}^{(1)}$, respectively. One possible observable which can characterize their correlation is the product $\cos \theta_{\rm open} = \hat{p}_{b} \cdot \hat{p}_{\bar{b}}$. The distribution of the variable can be written as
\begin{equation}
\frac{d\sigma}{d\cos\theta_{\rm open}}=1+D\cos\theta_{\rm open},
\end{equation}
where $D$ is a constant. The opening angle distribution is shown in Fig.~\ref{fig:jj_open} for both Monte Carlo truth and the reconstructed one. As expected, we obtain a flat distribution for SUSY and a slope for UED from the Monte Carlo truth, but the reconstructed distributions are again modified by experimental cuts and wrong combinations/solutions. The constant is $D$ is positive, indicating that the $b$-jets tend to move in the same direction. This can be understood as follows: the KK-bottom pair is more copiously produced from the gluon-gluon initial state than from $q\bar q$. Near the threshold, the final state has no orbital angular momentum, therefore has a total angular momentum of 0 or 1. The initial gluons do not have orbital angular momentum either and cannot form a spin-1 state.  Therefore the final state must have total spin 0, corresponding to $b^{(1)}$ and $\bar b^{(1)}$ of the same helicity. Due to the chiral coupling between $b$ and $b^{(1)}$, the resulting $b$ and $\bar b$ tend to go in the same direction. That being said, we note that the $q\bar q$ initial states contribute significantly to the total cross-section (37\%, comparing with 15\% for $t\bar t$) and give an opposite slope, which dilutes the effect and  makes it a difficult measurement. One can also look at the opening angle between the leptons. However, unlike the $t\bar t$ case where the charged lepton has the best distinguishing power, in our case, the charged lepton gives no advantage due to the fact that the charge of the sbottom is not correlated with the charge of the near lepton.  

One can combine the measurements of the production angle and jet-jet opening angle to optimize the distinguishing power. For example, we can define a ``central-forward'' asymmetry for the production angle
\begin{equation}
A_{\rm prod}=\frac{\sigma(|\cos\theta_{\rm prod}|>0.5) - \sigma(|\cos\theta_{\rm prod}|<0.5)}{\sigma_{\rm total}},
\end{equation}
and a ``forward-backward'' asymmetry for the jet-jet opening angle.  
\begin{equation}
A_{jj}=\frac{\sigma(\cos\theta_{\rm open}>0) - \sigma(\cos\theta_{\rm open}<0)}{\sigma_{\rm total}}.
\end{equation}
The expectation values for the asymmetries after event reconstruction are given by 
\begin{eqnarray}
&&A_{\rm prod} (\mbox{SUSY}) = 0.042\, (0.078), \  \ A_{\rm prod}(\mbox{UED})=0.119\, (0.164); \nonumber \\
&&A_{jj}(\mbox{SUSY})=0.003\, (0.000), \  \ A_{jj}(\mbox{UED}) = 0.026\,(0.023),
\end{eqnarray}
where the numbers in the parentheses are from the Monte Carlo truth for comparison.
The asymmetries are small, in which case the statistic errors are simply given by $1/\sqrt{N}$ where $N$ is the number of available events after cuts. By combining the two measurements, it is possible to distinguish the two spins with $\sim 1700$ events at 95\% level after cuts\footnote{Ignoring a small correlation between $A_{\rm prod}$ and $A_{jj}$, we define $\chi^2(\rm SUSY) = [A_{\rm prod}({\rm exp})-A_{\rm prod}({\rm SUSY})]^2/\sigma^2_{\rm prod}+ [A_{jj}({\rm exp})-A_{jj}({\rm SUSY})]^2/\sigma^2_{jj}$ and $\chi^2(\rm UED) = [A_{\rm prod}({\rm exp})-A_{\rm prod}({\rm UED})]^2/\sigma^2_{\rm prod}+ [A_{jj}({\rm exp})-A_{jj}({\rm UED})]^2/\sigma^2_{jj}$, where $A_{\rm prod}({\rm exp})$ and $A_{\rm jj}({\rm exp})$ are the experimental values and $\sigma_{\rm prod} = \sigma_{jj} = 1/\sqrt{N}$. We estimate the needed number of events by requiring that the  correct theory  has 95\% probability of having the smaller $\chi^2$ and hence being selected.}. This number is obtained by assuming a sample of pure signal events. In reality, more events may be required due to systematic uncertainties, the SM backgrounds, as well as contaminations from other new physics processes with the same final state particles. Of course, one should combine other information from the decay chain to better determine the models. For example, the jet-near lepton invariant mass gives us information about the spin of the $\cntwo$/KK-$Z_\mu$. For $\cntwo$, the distribution is flat since we do not know the charge of the $b$-jets, while for KK-$Z_\mu$ a second order polynomial can be seen~\cite{Smillie:2005ar, Athanasiou:2006ef}. However, we emphasize again that this kind of information is not a direct measurement of the spin of the first particle in the decay chain.

\section{Discussion and Conclusions}
\label{sec:conclusions}

In many scenarios of new physics beyond the Standard Model, such as supersymmetry and UED, the decay of new physics particles frequently leads to long decay chains ending with a stable massive neutral particle with undetectable momentum.  In the article, we have studied methods of reconstructing the kinematics of such decay chains. We began with the assumption that the masses of the new physics particles involved in the decay chain have been measured. We then showed that their momenta, in particular the momentum of the stable neutral particle, can be fully reconstructed. As an application of this method, we used the kinematic information to determine the spin of new particles and showed that different new physics scenarios, such as supersymmetry and UED, can be distinguished with this method. Well studied methods using Lorentz invariant variables are not directly applicable to the measurement of the spin of the particle at the first or the last step of the decay chain. With full kinematic information of the decay chain, we are able to probe the spin of the particle which initiates the decay chain. 

We performed two case studies. First, we considered the kinematic reconstruction and spin measurement with information from only one side of the event, {\it i.e.,} one decay chain. We also demonstrated a ``double chain" analysis, using the kinematic information to obtain the production angle and the correlation between the decay products of two new physics particles, one on each side of the events. We expect these two methods to be complementary. The final state in the single chain analysis is obvious more inclusive. At the same time, for the distribution of the decay products to contain useful spin information, the particle under consideration needs to be produced in a polarized state, and its coupling to its decay product has to be chiral. On the other hand, extraction of non-trivial spin correlation in the double chain analysis can be successful without such special requirements on the couplings of new physics states. However, it obviously requires precise knowledge of both sides of the decay chains. 

We have demonstrated our method using a set of particular benchmark models, SPS1a with a light gravitino and a model with similar mass spectrum in the cases of supersymmetry and UED, respectively. While these benchmark models are not designed to allow an easy spin measurement, our choices of the production channel and the strategy do take advantage of specific features of the spectrum. For example, we have relied on the fact that squark/KK-quark exchange does not contribute significantly  to the $\tilde{\chi}_2^0 \tilde{\chi}^+$/$Z^{(1)} W^{+(1)}$ production. Similar method should be applicable to other models, even though the specific choice of channels and strategy can be different. We remark that this situation is expected to be quite generic. Due to subtleties in the extraction of spin information and the virtually infinite number of possibilities of new physics models, it is impossible to have an observable which is universally applicable. However, once enough details of the new physics states, such as masses and quantum numbers,  are known, it is likely that specific variations of several proposed  classes of spin measurement methods, such as the method demonstrated here and the invariant mass method, can be adapted to accomplish the task. 

To focus on our demonstration of the reconstruction and spin measurement method, we have used exclusive samples of signal events in our analysis.  Realistically, achieving an exclusive sample of high purity requires strict cuts to suppress the Standard Model backgrounds, and contaminations from other new physics channels. Performing a careful study of the reach in the specific examples discussed here could be interesting. However, the design and optimization of such cuts will inevitably be very model dependent. As a result,  the conclusion of such a study is less likely to be representative in a large class of models. Therefore, such studies will be more effective after particular new physics channels have been identified at the LHC. Due to the expected low efficiency in isolating such exclusive samples, and the prerequisite of mass measurements, we expect the method presented here will be useful only with large statistics. 

Instead of assuming prior knowledge of the masses, we could in principle perform a combined strategy which fits both masses and spins. Moreover, reconstruction of momenta can also help measure other properties of the new physics, such as the chirality of the couplings as we have already alluded to in Section~\ref{sec:single}.  We will postpone further development of our methods in those directions to future studies. Finally, we note that momentum reconstruction is also useful for observing CP violations \cite{MoortgatPick:2009jy}. It is interesting to study possible applications of the methods presented in this article.

\acknowledgments  
The work of H.-C.  C. is supported in part by the Department of Energy Grant DE-FG02-91ER40674. Z.H. is supported in part by the National Science Foundation under grant PHY-0804450. I. W. K., is supported by the U.S. Department of Energy grant DE-FG-02-95ER40896. 
L.-T. W.~is supported by the National 
Science Foundation under grant PHY-0756966 and the Department of
Energy Outstanding Junior Investigator award under grant DE-FG02-90ER40542.
  
\appendix 
\section{$\chi^2$ Minimization Method for Momentum Reconstruction 
of Over-constrained Systems}
\label{app:likelihood} 

We describe the general formulation of momentum 
reconstruction for over-constrained systems such as the three-step 
cascade decay chains depicted in Fig.~\ref{fig:double_chain}.
We also point out several subtleties that arise from our assumptions about the measurement variables and address how to resolve these subtleties. 

The goal of event reconstruction is to determine the invisible
particle momenta that maximize the likelihood (by minimizing $\chi^2$) of hypothetical 
mass shell relations of the involved particles. 
The ingredients can be summarized as follows:
 \begin{itemize}
\item {\it Parameters}:  The parameters are the quantities to be reconstructed.  They are varied in the reconstruction procedure.  Here, the parameters are the momenta of the invisible particles, which we denote as
$\theta_I$, $I = 1,\dots, N$.

\item {\it Measurement variables}:   The measurement variables are fixed quantities for each event, such as the momenta of the visible particles (jets and leptons).  We denote these quantities by  ${\mathbf x}_k$,  $k = 1, \dots, m$.

\item {\it Nuisance parameters}:  These quantities are shared by all events and are assumed to be known {\it a priori}.    Here, the nuisance parameters are the mass parameters of the particles involved in the event topology. 
For simplicity, in our analysis we will treat these parameters as measurement variables ({\it i.e.}, as part of the ${\mathbf x}_k$'s). 

\item {\it Hypotheses}:  The hypotheses are the relations that should be satisfied by the parameters, measurement variables, and nuisance parameters of the system.  Here, these are the mass-shell relations and the
missing momentum sum.   We will label the hypotheses as
${\mathbf y}_i ({\mathbf x}_k ;  
                   \theta_I) = 0$, $i = 1, \dots, n$.
\end{itemize}

The procedure is to calculate the likelihood of the hypotheses and maximize the likelihood 
 with respect to the parameters. 
Assuming Gaussian statistics, the maximum likelihood condition
can be equivalently obtained by minimizing $\chi^2$ of the hypotheses for a
given event. 
Under this assumption, the statistical dependence between the measurement 
 variables can be described by the covariance matrix: 
\begin{eqnarray}
U_{kl}= \langle ({\mathbf x}_k - \overline{{\mathbf x}_k}) 
({\mathbf x}_l - \overline{{\mathbf x}_l}) \rangle.
\end{eqnarray} 
Here $\langle q \rangle$ and $\overline{q}$ denote the statistical mean 
value of some arbitrary variable $q$, evaluated over statistical ensemble of experimental measurements
with the same physical configuration.

To define the $\chi^2$ of the hypotheses $\{{\mathbf y}_i =0\}$, one defines the covariance matrix $V_{ij}$ of ${\mathbf y}_i$  at a given $({\mathbf x}_k, \theta_I)$ by
\begin{eqnarray}
V_{ij} &=&\langle  {\mathbf y}_i  {\mathbf y}_j  \rangle \\
&=& \sum_{k,l} 
\left. 
\frac{\partial {\mathbf y}_i}{\partial {\mathbf x}_k} 
\frac{\partial {\mathbf y}_j}{\partial {\mathbf x}_l}
\right|_{{\mathbf x},\theta} U_{kl},
\label{eq:VijandUkl}
\end{eqnarray}
where we have used the fact that ${\mathbf y}_i$'s must have zero mean value.  Since the covariance matrix $V_{ij}$ is not diagonal in the systems under consideration, the $\chi^2$ function of our hypotheses is given by 
\begin{eqnarray}
\chi^2 (\theta_I) = [{\mathbf y}({\mathbf x},\theta)]^T \cdot V^{-1} \cdot [{\mathbf y}({\mathbf x},\theta)], 
\label{eq:chisqr} 
\end{eqnarray} 
where $[{\mathbf y}]$ denotes a column vector constructed from the ${\mathbf y}_i$. 

For the cascade decays depicted in Fig.~\ref{fig:double_chain}, 
the kinematic constraints of the system can be summarized by the following set of 
equations:
\begin{eqnarray}
{\mathbf y}_{1} &=& p_1^2  - m^2_{N} = 0,  \\ 
{\mathbf y}_{2} &=& \left(p_1 + p_3 \right)^2  - m^2_X = 0,  \\ 
{\mathbf y}_{3} &=& \left(p_1 + p_3 + p_5 \right)^2  - m^2_{Y} = 0,  \\ 
{\mathbf y}_{4} &=& 
\left(p_1 + p_3 + p_5 + p_7  \right)^2  - m^2_{Z} = 0,  \\ 
{\mathbf y}_{5} &=& p_2^2  - m^2_{N} = 0,  \\ 
{\mathbf y}_{6} &=& \left( p_2 + p_4 \right)^2  - m^2_{X} = 0,  \\ 
{\mathbf y}_{7} &=& 
\left( p_2 + p_4 + p_6 \right)^2  - m^2_{Y} = 0,  \\ 
{\mathbf y}_{8} &=& 
\left( p_2 + p_4 + p_6 + p_8 \right)^2  - m^2_{Z} = 0,  \\ 
{\mathbf y}_{9} &=& p_1^{x} + p_2^{x} - \slashchar{p}^x = 0, \\
{\mathbf y}_{10} &=& p_1^{y} + p_2^{y} - \slashchar{p}^y = 0.
\end{eqnarray}
For this system, the $\theta_I$ parameters are the momentum variables for the invisible particles in each event, which are the eight real quantities
$p_1^{\mu}$ and 
$p_2^{\mu}$  ($\mu = 0,\dots,3$).  There are 30 measurement variables $\mathbf{x}_k$ for each event, which include the visible particle momenta  
$p_3^\mu, p_5^\mu, p_7^\mu, p_4^\mu, p_6^\mu, p_8^\mu$, the missing transverse momentum
$\slashchar{p}^x$ and $\slashchar{p}^y$, 
and the 4 ``nuisance parameters":
$m_{N}$, $m_{X}$,  $m_{Y}$ and $m_{Z}$.  The visible particle momenta are measured in terms of the transverse energy $E_T\equiv\sqrt{p_x^2+p_y^2+m^2}$, the pseudo-rapidity $\eta$ and the azimuthal angle $\phi$.  Since the visible particles in this case are massless, we can write
\begin{eqnarray}
p^\mu = (E_T c_\eta, E_T c_\phi, E_T s_\phi, E_T s_\eta),
\end{eqnarray}
where $c_\phi=\cos\phi$, $s_\phi=\sin\phi$, $c_\eta=\cosh\eta$, and $s_\eta=\sinh\eta$.
Denoting the statistical errors by $\delta_{E_T}, \delta_\eta$ and 
$\delta_\phi$ for $E_T, \eta$ and $\phi$, respectively, the covariance
matrix is given by 
\begin{eqnarray}
&& \langle p_\mu p_\nu \rangle = \\ 
&&\left(
\begin{array}{cccc}
\delta_{E_T}^2 c^2_\eta + E_T^2 \delta_\eta^2 s^2_\eta   &
\delta_{E_T}^2 c_\eta c_\phi                             &
\delta_{E_T}^2 c_\eta s_\phi                             &
(\delta_{E_T}^2 + E_T^2 \delta_\eta^2) c_\eta s_\eta       \\
\delta_{E_T}^2 c_\eta c_\phi                             & 
\delta_{E_T}^2 c^2_\phi + E_T^2 \delta_\phi^2 s^2_\phi   &
 (\delta_{E_T}^2 - E_T^2 \delta_\phi^2)  s_\phi c_\phi   &
\delta_{E_T}^2 s_\eta c_\phi                               \\
\delta_{E_T}^2 c_\eta s_\phi                             &
( \delta_{E_T}^2 - E_T^2 \delta_\phi^2)  c_\phi s_\phi   &
\delta_{E_T}^2 s^2_\phi + E_T^2 \delta_\phi^2 c^2_\phi   &
\delta_{E_T}^2 s_\eta s_\phi                               \\
(\delta_{E_T}^2 + E_T^2 \delta_\eta^2) s_\eta c_\eta     &
\delta_{E_T}^2 s_\eta c_\phi                             &
\delta_{E_T}^2 s_\eta s_\phi                             & 
\delta_{E_T}^2 s^2_\eta + E_T^2 \delta_\eta^2 c^2_\eta 
\end{array} 
\right). \nonumber
\end{eqnarray}
Here we have omitted the particle index for the same particle.  For different particles $i$ and $j$ we have $\langle p_\mu^i p_\nu^j \rangle = 0$, because the momentum measurements are uncorrelated for different particles. On the other hand, the missing transverse momentum measurement is correlated with the other measurements. 
The entries  of the covariance matrix involving $\slashchar{p}^{x}$ and $\slashchar{p}^{y}$ are thus given by
\begin{equation}
\langle \slashchar{p}_\mu p_\nu^i \rangle =-\langle {p}_\mu^i{p}_\nu^i\rangle, \ \ \langle \slashchar{p}_\mu \slashchar{p}_\nu \rangle = \sum_i \langle {p}_\mu^i{p}_\nu^i\rangle, \label{eq:ptcorr}
\end{equation}
where $\mu, \nu$ in Eqs~(\ref{eq:ptcorr}) are restricted to  $x$ and $y$.  The experimental errors from the measurements of particle momenta and missing transverse momentum are summarized in Table \ref{app:detector}. 
\begin{table}
\centering
\begin{tabular}{|cc|}
\hline
\multicolumn{2}{|l|}{Electrons and muons \cite{Ragusa:2007zz,cms_tdr}:}\\
\hline
 & $|\eta|< 2.4,\ \ p_T>10,$ \\
 & $\frac{\delta p_T}{p_T}=0.008\oplus 0.00015\,p_T,$ \\
 &  $\delta\theta = 0.001, \delta\phi = 0.001.$  \\
\hline
\multicolumn{2}{|l|}{ Photons \cite{cms_tdr,cms_ecal}: } \\
\hline
& $|\eta|< 3.0,\ \ p_T>10,$\\
& $\frac{\delta E}{E}=\frac{0.028}{\sqrt{E}}\oplus\frac{0.12}{E}\oplus0.0026,$ \\
& $\delta\eta = 0.0011,\ \ \delta\phi= 0.003.$ \\
\hline
\multicolumn{2}{|l|}{Jets: } \\
\hline
&$|\eta|< 3.0,\ \ p_T>100\GeV,$\\
&$\frac{\delta_{E_T}}{E_T}=\left\{
    \begin{array}{cl}\frac{5.6}{E_T}\oplus\frac{1.25}{\sqrt{E_T}}\oplus0.033, &\mbox{for }|\eta|<1.4,\\
                      \frac{4.8}{E_T}\oplus\frac{0.89}{\sqrt{E_T}}\oplus0.043, &\mbox{for }1.4<|\eta|<3.0, \\
                     \end{array}\right.$\\
&$\delta\eta=0.03, \ \ \delta\phi=0.02 \ \ \mbox{for }|\eta|<1.4,$\\
&$\delta\eta=0.02, \ \ \delta\phi=0.01\ \ \mbox{for }1.4<|\eta|<3.0.$\\
\hline
\end{tabular}
\caption{  Experimental errors from the measurements of particle momenta: In 
our analysis, parton level events are smeared according to the above Gaussian 
errors. The observables of energy dimension are in GeV units and the angular 
and the rapidity variables are in radians. 
Simple acceptance cuts on pseudo-rapidity $\eta$ are also applied.
For electrons and muons,
the resolution here corresponds roughly to the CMS tracking system performance in the central region ($\eta=0$) \cite{Ragusa:2007zz}. The resolution becomes slightly worse at higher rapidity until $|\eta|\gtrsim2$ where it starts to diverge. We ignore this rapidity dependent effect. 
For photons, the resolution for position measurement corresponds to the CMS ECAL performance obtained using electron beams ($10<p_T<50\GeV$) in Ref.~\cite{cms_ecal}.  
}
\label{app:detector}
\end{table}

With these ingredients, the next step is to determine $\chi^2$ as given in Eq.~(\ref{eq:chisqr}).  
However, one immediately encounters a difficulty in doing so because the covariant matrix $V$ is singular and the $\chi^2$ from Eq. (\ref{eq:chisqr}) is ill-defined. The physical reason for this singularity is that the two decay 
chains in Fig. \ref{fig:double_chain} are assumed to be symmetric and the masses of the particles in both chains are identical.  In a realistic situation, the assumption of symmetric chains can be incorrect due to effects such as finite decay widths and particle misidentification.   Nevertheless, it is desirable to have a method to resolve this problem without abandoning the symmetric chain assumption.   

Our procedure for addressing the situation is to introduce a regulator that controls the divergence as follows.  We double the number of nuisance parameters by assuming that the masses of the particles $N,\,X,\, Y,\, Z$ in each chain are effectively independent variables,  which leads to a covariance matrix e.g. for $m_N$ of the form 
\begin{eqnarray}
\langle (m_{Ni}-\overline{m_{Ni}}) (m_{Nj} - \overline{m_{Nj}}) \rangle 
= 
(\delta m_N)^2
\left(
\begin{array}{cc}
  1 & 1 \\
  1 & 1 
\end{array}   
\right).  \label{eq:masscov}
\end{eqnarray}
We then apply a very small perturbation $\epsilon$ to Eq. (\ref{eq:masscov}), 
\begin{eqnarray}
\langle (m_{Ni} - \overline{m_{Ni}}) (m_{Nj} - \overline{m_{Nj}}) 
\rangle = 
(\delta m_N)^2 
\left(
\begin{array}{cc}
  1 &  1+\epsilon \\
 1+\epsilon & 1 
\end{array}
\right).
\end{eqnarray} 
After this perturbation, $V_{ij}$ 
is no longer singular and $\chi^2$ is well-defined. The numerical results presented in Section~\ref{sec:double} are obtained by setting $\epsilon = 0.01$.
The divergent part of $\chi^2$ can be extracted by taking
$\epsilon  \rightarrow 0$,
\begin{eqnarray}
\chi^2 (\theta) = \widehat{\chi^2} (\theta) + \frac{1}{\epsilon} h(\theta).
\label{chisqcomp}
\end{eqnarray} 
The regular part of $\chi^2(\theta)$, which has been denoted as $\widehat{\chi^2}(\theta)$, can be extracted from the pseudoinverse of the symmetric matrix $V_{ij}$.  The pseudoinverse $\widetilde{V}$ is defined by  
\begin{eqnarray}
\widetilde{V} &=& W \cdot \widetilde{D} \cdot W^{-1}, 
\end{eqnarray}
where $W$ is the diagonalizing matrix of $V$: 
 \begin{eqnarray}
V = W \cdot D \cdot W^{-1} \mbox{,  with } D = {\rm diag} (d_1,d_2\dots d_m,0,0,0 \dots 0),\, (d_i \neq 0),
\end{eqnarray}
and $\widetilde{D}$ is given by
\begin{eqnarray}
{\widetilde{D}} &=& {\rm diag} (d_1^{-1},d_2^{-1},...,d_m^{-1},0,.. ).  
\end{eqnarray}
The regular part of $\chi^2$ is then given by 
\begin{eqnarray}
\widehat{\chi^2} = [{\mathbf y}]^T \cdot \widetilde {V} \cdot 
[{\mathbf y}].
\end{eqnarray}
The $1/\epsilon$ term plays the role of a penalty term since it effectively confines the 
configuration space (parameter space) to the solution space 
of $h(\theta) = 0$.
To obtain this term, we define the projection matrix $P$ of $V_{ij}$ onto the space of zero 
eigenvalues as follows:
\begin{eqnarray}
P (V)= W \cdot {\rm diag} ( 0,0,....0, 1,1,1,..) \cdot W^{-1},
\end{eqnarray}
where the 1's correspond to the $(n-m)$-dimensional subspace of $V$ with zero eigenvalues.  $h(\theta)$ in Eq.~(\ref{chisqcomp}) is then given by
\begin{eqnarray}
h(\theta)=[{\mathbf y}]^T \cdot P(V) \cdot [{\mathbf y}] =\sum_{k=m+1}^n \left (\sum_{i=1}^n \mathbf{y}_i W_{ik} \right)^2.
\end{eqnarray}
In summary, the $\chi^2$ minimization must be carried out in the restricted space of 
\begin{eqnarray}
\sum_{i=1}^n \mathbf{y}_i W_{ik} =0.
\end{eqnarray}

\end{document}